\documentclass[prd,aps,10pt,twocolumn,nofootinbib,showpacs]{revtex4-2}

\usepackage{amssymb,amsmath}
\usepackage{xcolor}
\usepackage[utf8]{inputenc}

\usepackage{upgreek}
\usepackage{graphicx}
\usepackage{framed}
\usepackage[makeroom]{cancel}

\usepackage[makeroom]{cancel}
\usepackage[caption=false]{subfig}
\usepackage[colorlinks=true,
            citecolor=green,
            linkcolor=red,
            filecolor=cyan,
            urlcolor=magenta,
            backref=false]{hyperref}

\newcommand{\dd}{\mbox{d}}

\begin{document}

\title{Torsion in two dimensions: autoparallels, symmetries, and applications to black holes}
\author{Jens Boos}
\email{jens.boos@kit.edu}
\affiliation{Institute for Theoretical Physics, Karlsruhe Institute of Technology, D-76128 Karlsruhe, Germany}

\date{April 8, 2025}

\begin{abstract}
We explore spacetime torsion in a two-dimensional setting, wherein it corresponds to a vector field. Without invoking field equations of a particular gravitational theory, we develop visualization techniques for such torsion fields, consider a generalization of Killing vectors to the presence of torsion, and find explicit representations of such Killing vectors (and their generalized algebra) in the presence of constant torsion. We then utilize these structures to derive general properties of surface gravities of static black holes whose near-horizon region features approximately constant torsion fields. Under certain assumptions, the two-dimensional results can be lifted to four dimensions.\\[10pt]
\textit{This paper is dedicated to the memory of Peter Baekler.}
\end{abstract}

\maketitle

\section{Introduction}

With tests of strong gravity reaching unprecedented levels of increasing accuracy, it becomes all the more important to check for blind spots. What are models of gravity that can presently not be tested with existing experiments? What are, nevertheless, promising modified gravity theories argued from first principles? In this paper, we would like to address this set of questions by focussing on post-Riemannian geometry with non-vanishing torsion, with a particular application to the near-horizon regions two-dimensional black holes as well as spherically symmetric four-dimensional black holes.

Although general relativity has passed all observational tests to date \cite{Will:2014kxa}, it is a fundamentally incomplete theory due to its perturbative nonrenormalizability. Quadratic gravity does not have this shortcoming, but instead features ghost-like instabilities. A possible solution to this problem lies in extending the geometric arena of gravity and allow for post-Riemannian quantities. In a minimal modification, gauging the Poincar\'e group in flat spacetime leads to Poincar\'e gauge gravity \cite{Hehl:1976kj}. Therein, a quadratic Lagrangian can be formulated in terms of the curvature as well as the torsion tensor, with a rich spectrum of theories allowing both for black hole solutions and different particle spectra, with some of them being ghost-free and stable \cite{Sezgin:1979zf,Sezgin:1981xs,Kuhfuss:1986rb,Karananas:2014pxa,Blagojevic:2018dpz,Percacci:2020ddy,Barker:2024ydb}.

However, the plethora of different choices of gravitational Lagrangians obfuscates the emergence of new, universal phenomena linked to the existence of spacetime torsion, in, say, black hole spacetimes. Moreover, while curvature as a concept has been well-established at the intuitive level, the same thing cannot be said about spacetime torsion. To that end, we will consider here a simplified scenario. First, we limit our considerations to two dimensions (and, later, suitably generalized to spherical symmetry in the four-dimensional sense).\footnote{Poincar\'e gauge gravities in two dimensions have been explored in \cite{Mielke:1993nc}, whereas the mathematical classification of two-dimensional spaces with torsion is fairly recent \cite{DAscanio:2019bei}; see also \cite{Agricola:2012eze}.} And second, for the later part we will consider constant torsion coefficients. And third, we consider such torsion fields in the presence of a flat metric.

Restricting as these simplifications may appear, we will demonstrate that they are sufficient to not only develop different visualization techniques for torsion, but also allow for the study of symmetries in post-Riemannian geometries, with universal consequences for the surface gravity of static, two-dimensional black holes, as well as for the surface gravity of a certain class of four-dimensional black holes.

This paper is organized as follows: in Sec.~\ref{sec:2} we introduce our notation and review basic definitions of spacetime torsion and related concepts. In Sec.~\ref{sec:3} we will introduce our model and discuss its geometric properties. Sec.~\ref{sec:4} is devoted to the visualization of generic torsion fields in two dimensions (both spacetime and purely spatial), before we then focus on the case of constant torsion in Sec.~\ref{sec:5}, including a description of its Killing vectors, suitably generalized. In Sec.~\ref{sec:6} we then apply the existence of the Killing vectors (i) to the near-horizon region of a two-dimensional static black hole and extract its modified surface gravity, and (ii) to the four-dimensional symmetry-reduced system of a static, spherically symmetric black hole. We summarize our findings in Sec.~\ref{sec:7}, and list relevant next steps and open problems.

\section{Preliminaries and notation}
\label{sec:2}
In a spacetime with curvature and torsion, the commutator of covariant derivatives acting on a vector field is related to the curvature tensor $R{}_{\mu\nu}{}^\rho{}_\sigma$ and the torsion tensor $T{}_{\mu\nu}{}^\lambda$ via
\begin{align}
\left[ \nabla{}_\mu, \nabla{}_\nu \right] V{}^\rho = R{}_{\mu\nu}{}^\rho{}_\alpha V{}^\alpha - T{}_{\mu\nu}{}^\alpha\nabla{}_\alpha V{}^\rho \, .
\end{align}
The torsion tensor is a vector-valued 2-form, $T{}_{\mu\nu}{}^\lambda = - T{}_{\nu\mu}{}^\lambda$, and, as such, has $\#(T) = n^2(n-1)/2$ independent components in $n$ spacetime dimensions. In the case of interest for this paper we have $n=2$, which equates to $\#(T) = 2$. This makes intuitive sense, since a 2-form in two dimensions is dual to a scalar function, and hence only the vector index contributes independent components. Lowering all indices on the curvature tensor, we find the symmetries
\begin{align}
R{}_{\mu\nu\rho\sigma} = -R{}_{\nu\mu\rho\sigma} \, , \quad
R{}_{\mu\nu\rho\sigma} = -R{}_{\mu\nu\sigma\rho}
\end{align}
resulting in $\#(R) = n^2(n-1)^2/4$. The symmetries are consistent with viewing the curvature tensor as a Lorentz-valued 2-form. Note that $R{}_{\mu\nu\rho\sigma} \not= R{}_{\mu\nu\rho\sigma}$, which is a purely Riemannian relation known from general relativity. In the case of $n=2$ we find $\#(R) = 1$, which again makes sense since the Lorentz group in two dimensions consists only of a single boost.

The above commutation relation is equivalent to the following definition of the covariant derivative of a $(1,1)$-tensor (and suitably generalized for others) via
\begin{align}
\label{eq:cov-derivative}
\nabla{}_\mu X{}^\nu{}_\rho &\equiv \partial{}_\mu X{}^\nu{}_\rho + \Gamma{}^\nu{}_{\mu\alpha} X{}^\alpha{}_\rho - \Gamma{}^\alpha{}_{\mu\rho} X{}^\nu{}_\alpha \, .
\end{align}
Here, $\Gamma{}^\lambda{}_{\mu\nu}$ denotes the connection, and it can be split into a Riemannian part $\widetilde{\Gamma}{}^\lambda{}_{\mu\nu}$ as well as a torsion contribution $K{}^\lambda{}_{\mu\nu}$ (called the contortion tensor) as follows:
\begin{align}
\Gamma{}^\lambda{}_{\mu\nu} &= \widetilde{\Gamma}{}^\lambda{}_{\mu\nu} + K{}^\lambda{}_{\mu\nu} \, , \\
\widetilde{\Gamma}{}^\lambda{}_{\mu\nu} &= \frac12 g{}^{\lambda\alpha} \left( \partial{}_\mu g{}_{\alpha\nu} + \partial{}_\nu g{}_{\alpha\mu} - \partial{}_\alpha g{}_{\mu\nu} \right) \, , \\
\label{eq:contortion}
K{}^\lambda{}_{\mu\nu} &= \frac12 \left( T{}_{\mu\nu}{}^\lambda - T{}_\mu{}^\lambda{}_\nu - T{}_\nu{}^\lambda{}_\mu \right)\, .
\end{align}
Useful symmetry relations are
\begin{align}
\Gamma{}^\lambda{}_{(\mu\nu)} &= \widetilde{\Gamma}{}^\lambda{}_{\mu\nu} - \frac12 \left( T{}_\mu{}^\lambda{}_\nu + T{}_\nu{}^\lambda{}_\mu \right) \, , \\
2\Gamma{}^\lambda{}_{[\mu\nu]} &= T{}_{\mu\nu}{}^\lambda \, , \\
K{}^\lambda{}_{(\mu\nu)} &= -\frac12 \left( T{}_\mu{}^\lambda{}_\nu + T{}_\nu{}^\lambda{}_\mu \right) \, , \\
K{}_{(\lambda\mu)\nu} &= 0 \, ,
\end{align}
where $(ab) \equiv (ab+ba)/2$ and $[ab] \equiv (ab-ba)/2$ are the normalized (anti-)symmetrizers. Using the above concepts, we arrive at the following expressions for the curvature tensor and torsion tensor, respectively:
\begin{align}
R{}_{\mu\nu}{}^\rho{}_\sigma &= \partial{}_\mu \Gamma{}^\rho{}_{\nu\sigma} + \Gamma{}^\rho{}_{\mu\alpha} \Gamma{}^\alpha{}_{\nu\sigma} - (\mu \leftrightarrow \nu) \, , \\
T{}_{\mu\nu}{}^\lambda &= \Gamma{}^\lambda{}_{\mu\nu} - \Gamma{}^\lambda{}_{\nu\mu} \, .
\end{align}
We emphasize that all of the above expressions are formulated with respect to a coordinate basis $x{}^\mu$.

We close this introductory section by reviewing the form of the geodesic equation of motion, the autoparallel equation of motion, as well as the shape of the Killing equation, since they form an integral part of the subsequent considerations.

Since the quantity $K{}^\lambda{}_{\mu\nu}$ is a tensor, both $\Gamma{}^\lambda{}_{\mu\nu}$ and $\widetilde{\Gamma}{}^\lambda{}_{\mu\nu}$ are well-defined connections. For this reason, we define the Riemannian covariant derivative as $\widetilde{\nabla}{}_\mu$ in exact analogy to Eq.~\eqref{eq:cov-derivative}, 
\begin{align}
\label{eq:cov-derivative-riemann}
\widetilde{\nabla}{}_\mu X{}^\nu{}_\rho &\equiv \partial{}_\mu X{}^\nu{}_\rho + \widetilde{\Gamma}{}^\nu{}_{\mu\alpha} X{}^\alpha{}_\rho - \widetilde{\Gamma}{}^\alpha{}_{\mu\rho} X{}^\nu{}_\alpha \, .
\end{align}
Considering a particle trajectory $x{}^\mu(\tau)$ and its 4-velocity $u{}^\mu = \partial_\tau x{}^\mu$, we can now define two equations of motion:
\begin{align}
& u{}^\alpha \nabla{}_\alpha u{}^\mu = 0 \quad \Leftrightarrow \quad \dot{u}{}^\mu + \left( \widetilde{\Gamma}{}^\mu{}_{\alpha\beta} - T{}_\alpha{}^\mu{}_\beta \right) u^\alpha u^\beta = 0 \, , \nonumber \\
& u{}^\alpha \widetilde{\nabla}{}_\alpha u{}^\mu = 0 \quad \Leftrightarrow \quad \dot{u}{}^\mu + \widetilde{\Gamma}{}^\mu{}_{\alpha\beta} u^\alpha u^\beta = 0 \, .
\end{align}
The first equation is called the \emph{autoparallel equation} whereas the second is the \emph{geodesic equation}. Clearly, in the case of vanishing torsion, the two notions coincide. Note that the normalization of 4-velocity is preserved in the autoparallel case since $T{}_{\alpha\beta\gamma}u{}^\alpha u{}^\beta u{}^\gamma = 0$.

Turning to the Killing equation of general relativity for a vector $\widetilde{\xi}{}^\mu$, utilizing the Lie derivative $\mathcal{L}$ one has
\begin{align}
\mathcal{L}_{\widetilde{\xi}} \, g{}_{\mu\nu} = 0 \quad \Leftrightarrow \quad \widetilde{\nabla}{}_\mu \widetilde{\xi}{}_\nu + \widetilde{\nabla}{}_\nu \widetilde{\xi}{}_\mu = 0 \, .
\end{align}
As is well known, this not only justifies the interpretation of $\widetilde{\xi}{}^\mu$ as an isometry, it also implies that the quantity
\begin{align}
\widetilde{Q} = g{}_{\alpha\beta}u{}^\alpha \widetilde{\xi}{}^\beta
\end{align}
is conserved along geodesic motion, $u{}^\alpha\widetilde{\nabla}{}_\alpha u{}^\mu = 0$. While it is possible to construct an analog of Killing vectors in the presence of torsion---motivated by isometry considerations---this is not the path we will follow here. Instead, similar to \cite{Peterson:2019uzn}, we define an \emph{autoparallel Killing vector} $\xi{}^\mu$ as a solution of the equation\footnote{For different notions of Killing vectors in the presence of torsion we refer to \cite{DAscanio:2019tpq,Obukhov:2022khx}.}
\begin{align}
\nabla{}_\mu \xi{}_\nu + \nabla{}_\nu \xi{}_\mu = 0 \, .
\end{align}
Given such a vector, the quantity
\begin{align}
Q = g{}_{\alpha\beta}u{}^\alpha \xi{}^\beta
\end{align}
is conserved along autoparallel motion, $u{}^\alpha\nabla{}_\alpha u{}^\mu = 0$, making this vector a useful construct in the study of autoparallel motion.

In order to interpret Killing vectors, it is helpful to compute their commutator. For two vectors $X{}^\mu$ and $Y{}^\mu$, their commutator is
\begin{align}
[X,Y]{}^\mu = X{}^\alpha \partial{}_\alpha Y{}^\mu - Y{}^\alpha \partial{}_\alpha X{}^\mu \, .
\end{align}
The result is again a vector. We can make this apparent by inserting instead $\widetilde{\nabla}{}_\mu$, arriving at
\begin{align}
[X,Y]{}^\mu = X{}^\alpha \widetilde{\nabla}{}_\alpha Y{}^\mu - Y{}^\alpha \widetilde{\nabla}{}_\alpha X{}^\mu \, .
\end{align}
In order to define a similar commutator in the presence of torsion, we define the \emph{T-commutator} between two vectors $X{}^\mu$ and $Y{}^\mu$ as follows:
\begin{align}
\begin{split}
[X,Y]{}^\mu_\text{T} &\equiv X{}^\alpha \nabla{}_\alpha Y{}^\mu - Y{}^\alpha \nabla{}_\alpha X{}^\mu \\
&= [X,Y]{}^\mu + T{}_{\alpha\beta}{}^\mu X{}^\alpha Y{}^\beta \, .
\end{split}
\end{align}
As in the previously introduced concepts, in the limiting case of vanishing torsion we recover the notions from general relativity: the T-commutator reduces to the ordinary commutator $[X,Y]{}^\mu$ in that limit. Last, let us emphasize that this commutator satisfies the usual Leibniz rules under multiplication with a scalar function $f$,
\begin{align}
\begin{split}
[f\,X,Y]_\text{T} &= f [X,Y]_\text{T} - (\partial_Y f) X \, , \\
[X,f\,Y]_\text{T} &= f [X,Y]_\text{T} + (\partial_X f) Y \, .
\end{split}
\end{align}

\section{Main setting}
\label{sec:3}
With the notational machinery developed, let us now define the two-dimensional geometry under consideration for the remainder of this paper. To that end, will consider the following metric and torsion coefficients:
\begin{align}
\label{eq:geometry}
\dd s^2 = -\epsilon \dd t^2 + \dd x^2 \, , \quad T{}_{tx}{}^t = \tau \, , \quad T{}_{tx}{}^x = \chi \, .
\end{align}
In the above, $t$ and $x$ take values in $\mathbb{R}$. Setting $\epsilon = +1$ allows us to study spacetime, and $\epsilon = -1$ instead renders the above a purely spatial metric (and in that case we will change variable names $t \rightarrow y$). Initially, we will consider $\tau = \tau(t,x)$ and $\chi = \chi(t,x)$ and will later focus on the case of constant torsion coefficients, $\tau = \text{const}$ and $\chi = \text{const}$. Here and in what follows, we will denote derivatives with respect to $t$ with a dot, and derivatives with respect to $x$ with a prime. For some function $f(t,x)$,
\begin{align}
\dot{f}(t,x) \equiv \partial_t f(t,x) \, , \quad f'(t,x) \equiv \partial_x f(t,x) \, .
\end{align}
In the two-dimensional flat case, in coordinates $t$ and $x$, one has $\widetilde{\Gamma}{}^\lambda{}_{\mu\nu} = 0$ such that $\Gamma{}^\lambda{}_{\mu\nu} = K{}^\lambda{}_{\mu\nu}$. The curvature tensor then is simply
\begin{align}
\label{eq:curvature-in-terms-of-torsion}
R{}_{\mu\nu}{}^\rho{}_\sigma \overset{\ast}{=} \partial{}_\mu K{}^\rho{}_{\nu\sigma} - \partial{}_\nu K{}^\rho{}_{\mu\sigma} \, ,
\end{align}
where the contribution quadratic in $K{}^\lambda{}_{\mu\nu}$ vanishes identically in two dimensions. Hence, constant torsion gives no contribution to the curvature. The curvature tensor in two dimensions, even in the presence of torsion, is uniquely specified by one component. Consistent with the previous statement, it is proportional to derivatives of the torsion tensor, and its single unique component is given by
\begin{align}
R{}_{tx}{}^t{}_x = -\tau' - \epsilon \dot{\chi} = \frac12 R \, ,
\end{align}
where $R = g{}^{\mu\nu} R{}_{\alpha\mu}{}^\alpha{}_\nu$ is the Ricci scalar. This consideration highlights the fact that even though the background metric \eqref{eq:geometry} is flat, the presence of torsion generates a non-trivial spacetime curvature. Mathematically, this is due to the semidirect product structure of the Poincar\'e group (rotations and translations do not commute, and therefore the translational field strength, torsion, ``talks'' to the rotational field strength, curvature).

For completeness, let us mention that the torsion square is given by
\begin{align}
T{}_{\mu\nu\rho}T{}^{\mu\nu\rho} = 2(\tau^2-\epsilon\chi^2) \, .
\end{align}
Hence, if $\tau = \chi$ and $\epsilon=1$ we arrive at the special case of \emph{null torsion} which we shall address separately later.

Let us close this introductory section by embedding the geometry \eqref{eq:geometry} in Poincar\'e gauge gravity, whose vacuum field equations in our conventions can be written as \cite{Obukhov:2022khx}
\begin{align}
& a_0 \left( R{}_{\mu\nu} - \frac12 R g{}_{\mu\nu} \right) + \Lambda g{}_{\mu\nu} + {q}^\text{T}_{\mu\nu} + \ell_\text{Pl}^2 \, q^\text{R}_{\mu\nu} \nonumber \\
& - (\nabla{}_\alpha + T{}_\alpha) h{}_\nu{}^\alpha{}_\mu - \frac12 T{}_{\alpha\beta\nu} h{}^{\alpha\beta}{}_\mu = \kappa T{}_{\mu\nu} \, , \\
& a_0( T{}_{\mu\nu}{}^\lambda + T{}_\mu \delta{}_\nu^\lambda - T{}_\nu \delta{}_\mu^\lambda ) - h{}^\lambda{}_{\mu\nu} + h{}^\lambda{}_{\nu\mu} \nonumber \\
&-2\ell_\text{Pl}^2\left[ (\nabla_\alpha - T{}_\alpha) h{}^{\lambda\alpha}{}_{\mu\nu} + \frac12 T{}_{\alpha\beta}{}^\lambda h{}^{\alpha\beta}{}_{\mu\nu} \right] = \kappa S{}_{\mu\nu}{}^\lambda \, , \nonumber
\end{align}
where $T{}_{\mu\nu}$ is the energy-momentum tensor, and $S{}_{\mu\nu}{}^\lambda$ is the spin-angular momentum tensor, and $T{}_\mu$ denotes the torsion trace covector,
\begin{align}\begin{split}
T{}_\mu \equiv T{}_{\alpha\mu}{}^\alpha = (-\chi, \tau) \, , \\
T{}^\mu = g{}^{\mu\alpha} T{}_\alpha = (\epsilon \chi , \tau) \, .
\end{split}
\end{align}
We moreover defined
\begin{align}
q{}^\text{T}_{\mu\nu} &= T{}_{\mu\alpha\beta} h{}{}_\nu{}^{\alpha\beta} - \frac14 g{}_{\mu\nu} T{}_{\alpha\beta}{}^\gamma h{}^{\alpha\beta}{}_\gamma \, , \\
h{}^{\mu\nu}{}_\lambda &= a T{}^{\mu\nu}{}_\lambda \, , \\
q{}^\text{R}_{\mu\nu} &= R{}_{\mu\alpha}{}^{\beta\gamma} h{}{}_\nu{}^\alpha{}_{\beta\gamma} - \frac14 g{}_{\mu\nu} R{}_{\alpha\beta}{}^{\gamma\delta} h{}^{\alpha\beta}{}_{\gamma\delta} \, , \\
h{}^{\mu\nu}{}_{\rho\sigma} &= b R{}^{\mu\nu}{}_{\rho\sigma} \, .
\end{align}
In general, the expressions $h{}^{\mu\nu}{}_\lambda$ and $h{}^{\mu\nu}{}_{\rho\sigma}$ can be further expanded into irreducible pieces of torsion and curvature, respectively, but we do not proceed along those lines here since we are interested in the two-dimensional case wherein torsion and curvature are already irreducible.

Using computer algebra is is straightforward to show that the above geometry, in the special case of constant torsion coefficients, corresponds to an exact solution of the field equations with constant energy-momentum and spin-angular momentum, who are proportional to the coupling constant $a$:
\begin{align}
T{}_{\mu\nu} &= -\frac{a}{2} \left( \begin{matrix}
\epsilon\tau_0^2+\chi_0^2 & 2\epsilon\tau_0\chi_0 \\
2\epsilon\tau_0\chi_0 & \tau_0^2+\epsilon\chi_0^2 
\end{matrix} \right) \, , \\
S{}_{\mu\nu}{}^\lambda &= -a \, \left(\begin{matrix}
0 & \left(\begin{matrix} \tau_0 \\ \chi_0 \end{matrix}\right) \\
(-1)\left(\begin{matrix} \tau_0 \\ \chi_0 \end{matrix}\right) & 0 
\end{matrix}\right) = -\frac{a}{2} \, \epsilon_{\mu\nu} \epsilon{}^{\alpha\beta}T{}_{\alpha\beta}{}^\lambda \, . \nonumber
\end{align}
In particular, one recognizes that in case of null torsion the local energy density vanishes, and for torsion purely in temporal or spatial direction (that is, if $\tau_0\chi_0=0$) the transverse pressures vanishes. Moreover, the spin density is proportional to the torsion tensor in this simple two-dimensional example of constant torsion.

If the torsion coefficients are general functions, it is more difficult to extract physical intuition related to the energy-momentum tensor as well as the spin density. This motivates a graphical analysis of various torsion fields, which will be the subject of the following section.

\section{Generic torsion: visualizations}
\label{sec:4}

Curvature, as a concept, can be visualized straightforwardly since curved spaces can often be embedded into Euclidean space. This is helpful in building physical intuition for the geometric picture of general relativity's description of gravitation. Conversely, spaces with non-vanishing torsion can often bot be embedded in Euclidean space. With increased focus on torsion both in teleparallel gravity \cite{Aldrovandi:2013wha} as well as the physics of solids \cite{Puntigam:1996vy,Roychowdhury:2017}, different strategies need have been employed for visualization techniques. In two dimensions, those techniques are particularly simple, and we will briefly review existing ones and then build on them slightly, with a particular focus on autoparallels and the operational understanding of torsion.

For the sake of concreteness, we would like to focus on three concrete examples of torsion backgrounds, two of which are highly symmetrical, and one of which is more generic:
\begin{enumerate}
\item Rotational torsion:
\begin{align}
\label{eq:ex1}
T_{tx}{}^\mu = (\tau, \chi) = 0.3(-x, t) 
\end{align}
\item Boost torsion:
\begin{align}
\label{eq:ex2}
T_{tx}{}^\mu = (\tau, \chi) = 0.3(x, t)
\end{align}
\item Generic torsion:
\begin{align}
\label{eq:ex3}
T_{tx}{}^\mu = (\tau, \chi) = 0.15(t^2-0.4x^2-1, t - 0.9x)
\end{align}
\end{enumerate}
To track the influence of the metric signature, we will consider all three torsion configurations for both the spacetime case ($\epsilon=+1$) as well as the spatial case ($\epsilon=-1$), resulting in a total of six distinct scenarios. Recall that in the spatial case we relabel $t \rightarrow y$ for convenience.

\subsection{Canonical visualization (local)}

\begin{figure}[!htb]
\centering
\includegraphics[width=0.25\textwidth]{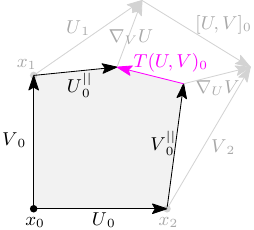}
\caption{We sketch the canonical visualization of an infinitesimal parallelogram (shaded grey) spanned by the vectors $U$ and $V$ at the point $x_0$. The subscripts on quantities denote whether they are to be evaluated at the point $x_0$ or the infinitesimally close points $x_1$ or $x_2$. The quantities $U_0^{||}$ and $V_0^{||}$ denote the parallel propagated vectors along each other. The abbreviation $T(U,V)$ stands for the vector $T{}_{\alpha\beta}{}^\mu U{}^\alpha V{}^\beta$. The grey arrows connect this infinitesimal parallelogram to the commutator of the two vector fields.}
\label{fig:canonical-sketch}
\end{figure}

We begin by recalling that torsion prevents infinitesimal parallelograms to close, as highlighted by Cartan (and summarized by 
Hehl \& Obukhov \cite{Hehl:2003,Hehl:2007bn}) or Schouten \cite{Schouten:1954}. Assuming that this parallelogram is spanned by two vectors $U{}^\mu$ and $V{}^\mu$, we may compute the parallel transport of them along each other. The change of $U{}^\mu$ parallel transported along $V{}^\mu$, to leading order, is given by
\begin{align}
\delta U{}^\mu = V{}^\alpha \partial{}_\alpha U{}^\mu &= - \Gamma{}^\mu{}_{\alpha\beta} V{}^\alpha U{}^\beta \overset{*}{=} - K{}^\mu{}_{\alpha\beta} V{}^\alpha U{}^\beta \, ,
\end{align}
where the last equality holds in the flat spacetime coordinates where $\widetilde{\Gamma}{}^\lambda{}_{\mu\nu} = 0$. Hence, the torsion tensor emerges as the difference of the two changes,
\begin{align}
\begin{split}
\delta U{}{}^\mu - \delta V{}^\mu &= - \Gamma{}^\mu{}_{\alpha\beta} (V{}^\alpha U{}^\beta - V{}^\beta U{}^\alpha) \\
&= T{}_{\alpha\beta}{}^\mu U{}^\alpha V{}^\beta \, .
\end{split}
\end{align}
This torsion projection is related to the commutator of two vector fields $U{}^\mu$ and $V{}^\mu$,
\begin{align}
\begin{split}
[U,V]{}^\mu_\text{T} &\equiv U{}^\alpha \nabla{}_\alpha V{}^\mu - V{}^\alpha \nabla{}_\alpha U{}^\mu \\
&= [U,V]{}^\mu + T{}_{\alpha\beta}{}^\mu U{}^\alpha V{}^\beta \, .
\end{split}
\end{align}
Evaluated in vicinity of a point $x_0$, as pointed out by Hehl \& Obukhov \cite{Hehl:2003,Hehl:2007bn}, this gives rise to what we shall refer to as the ``canonical'' visualization of torsion in the presence of two fiducial vector field $U{}^\mu$ and $V{}^\mu$. This visualization is local in nature, and is depicted schematically in Fig.~\ref{fig:canonical-sketch}.

\begin{figure*}[!htb]
\centering
\includegraphics[width=0.32\textwidth]{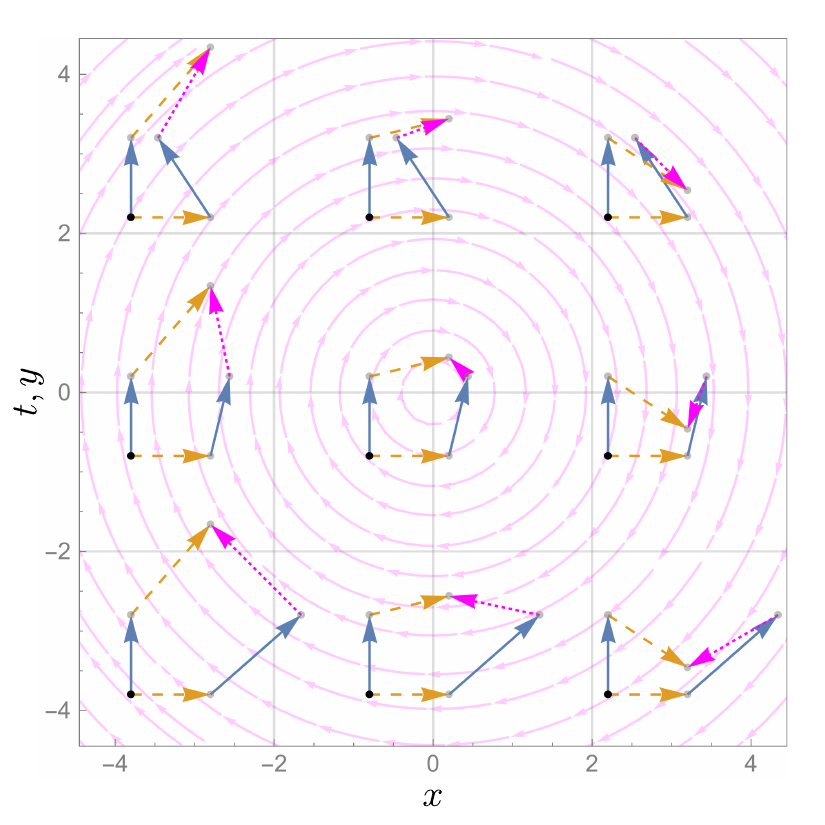}
\includegraphics[width=0.32\textwidth]{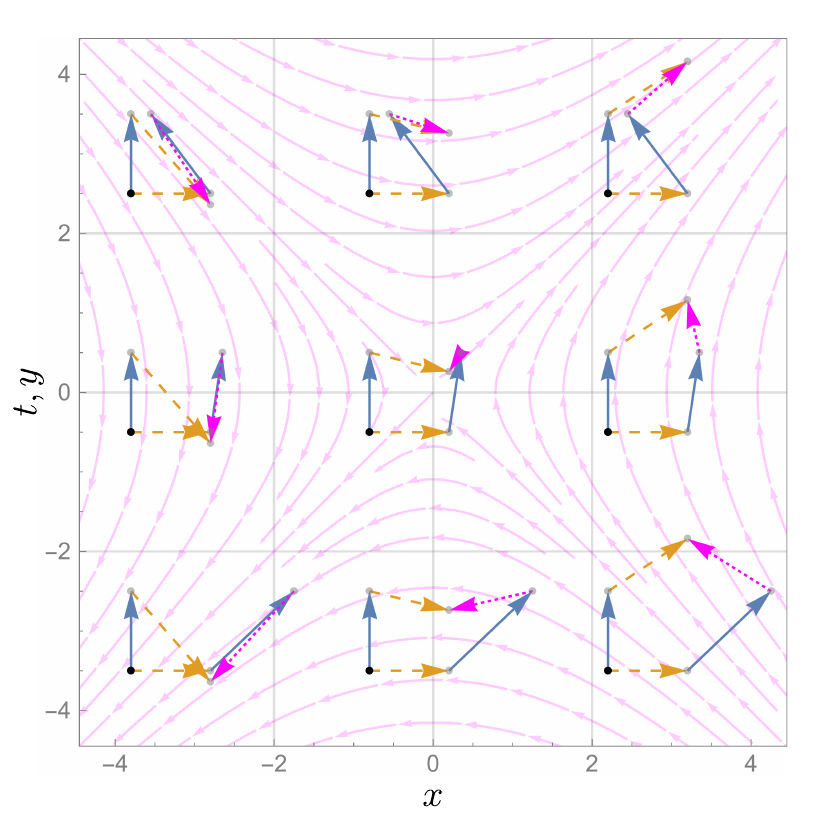}
\includegraphics[width=0.32\textwidth]{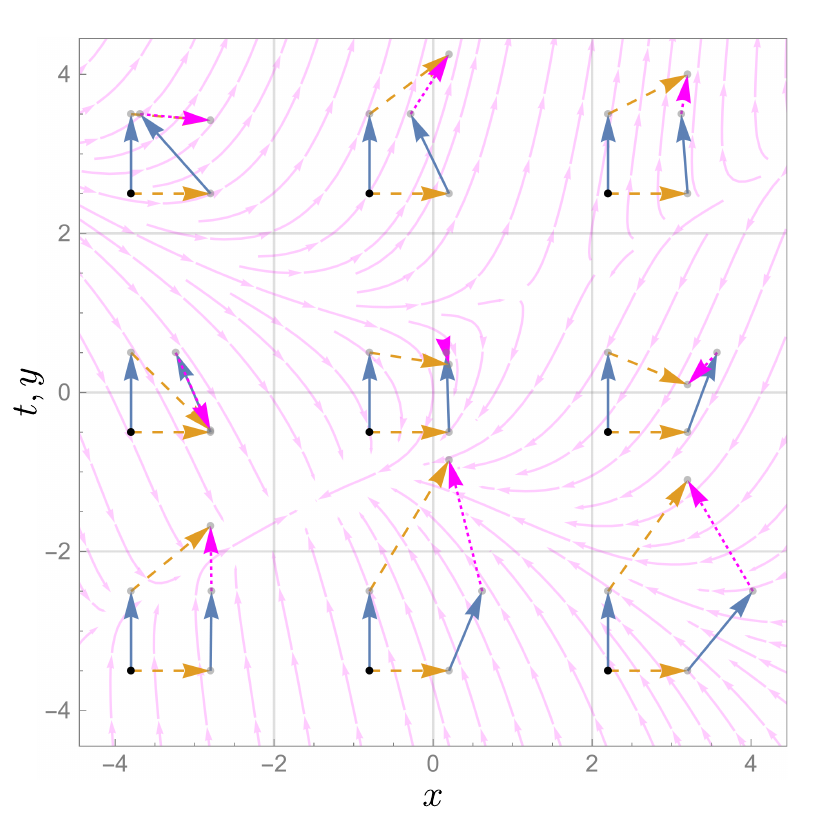}
\caption{We plot the canonical visualization of the three torsion fields \eqref{eq:ex1}, \eqref{eq:ex2}, and \eqref{eq:ex3}. The background streamplot shows the torsion field, and we implemented nine samples of the canonical visualization at various points throughout the plots. Due to the geometric nature of this visualization that does not make any reference to spacetime signature, the diagrams are identical for both the spacetime case ($\epsilon=+1$) and the purely spatial case ($\epsilon = -1$). At each sample point, the straight arrows represent the fiducial vector field $U = (1,0)$ and its parallel displaced copy along the other fiducial field $V = (0,1)$; analogously, the dashed arrow corresponds to the second fiducial field $V$ and its parallel displaced copy along $U$. Finally, the dotted arrow visualizes the torsion tensor projected to the two vector fields at the given point. Observe that this vector is parallel to the streamplot at each observation point (highlighted in bold).}
\label{fig:canonical-plots}
\end{figure*}

Now we are ready to apply the canonical visualization to our examples \eqref{eq:ex1}--\eqref{eq:ex3}, where for simplicity we will focus on the two fiducial vector fields
\begin{align}
U{}^\mu = (1,0) \, , \quad V{}^\mu = (0, 1) \, .
\end{align}
However, we emphasize that this visualization holds for any choice of vector fields, but the visual correspondence of $T{}_{\alpha\beta}{}^\mu U{}^\alpha V{}^\beta$ to the torsion background field $T{}_{tx}{}^\mu$ will be somewhat obfuscated. That being said, the results for the above choice is shown in Fig.~\ref{fig:canonical-plots}.

\subsection{Holonomy visualization (semi-local)}

While the canonical visualization is convenient and easy to implement (since it only utilizes local information) this is precisely its shortcoming. Namely, it does not make any statements about properties beyond the local limit. An important object in geometries with torsion are autoparallel curves that parallel propagate their own tangent vector. For this reason, we develop the above visualization further and define spacetime or spatial ``circuit'' that is constructed from two fiducial vector fields $U$ and $V$. The presence of torsion, as we will see, will prevent this circuit from being closed.

We begin with an autoparallel curve starting at a point $x_0$ whose initial tangent vector is $U$ and follow this autoparallel for a certain amount of proper time $\delta\lambda$, arriving at a point $x_1$. There, we define another autoparallel starting instead with the velocity $V$. Arriving at $x_2$ after some proper time $\delta\lambda$, we return along an autoparallel with velocity $-U$ to arrive at point $x_3$, and finally, for the last leg of the journey, we leave that point along direction $-U$ for another amount of proper time $\delta\lambda$. 

In order to avoid acausality in the spacetime case, we define the circuit via four timelike vectors, two of which are past-directed and two of which are future-directed. As a result, one may think of starting at point $x_2$ and sending two future-directed timelike autoparallels that end up at points $x_0$ and $x_4$. This ensures that an operational understanding of this measurement of torsion can be given in a straightforward manner: two observers, following first the vector field $U$ and then the vector field $V$ for a fixed amount of proper time (and the other way around for the second observer), will not meet at the same place if the torsion of space(time) is nonvanishing.

\begin{figure}[!htb]
\centering
\includegraphics[width=0.45\textwidth]{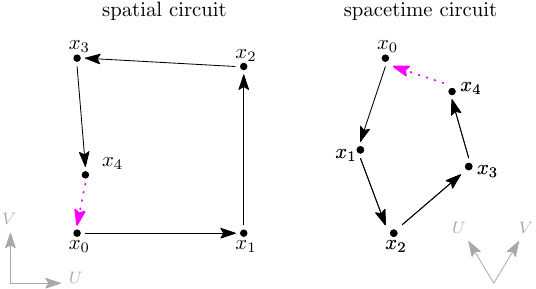}
\caption{We sketch the circuits utilized in the holonomy visualization of torsion based on the two fiducial vector fields $U$ and $V$. In the spatial case, we utilize two vectors normalized to $U \cdot U = v \cdot V = +1$, simply given by $U=(0,1)$ and $V=(1,0)$. In the spacetime case, however, we make use of timelike vectors, $U \cdot U = V \cdot V = -1$ with the parametrizations $U = (\sqrt{1 + v^2}, -v)$ and $V = (\sqrt{1+v^2}, +v)$, where $v \in (0,1)$ is the coordinate velocity of the observers. The distance between the points $x_0$ and $x_4$ is related to the presence of torsion.}
\label{fig:holonomy-sketch}
\end{figure}

Since in both cases we are travelling around an almost closed loop, a straightforward visualization consists of plotting the four autoparallels as well as the defect vector between the two points $x_0$ and $x_4$. Since this method is semi-local it will not give an exact result, but we will see that for small amounts of proper time $\delta\lambda$ the defect vector will be a simple linear combination of torsion coefficients. Before computing their theoretical form to leading order, see Fig.~\ref{fig:holonomy-sketch} for a visualization of the spatial circuit as well as the spacetime circuit utilized in the ``holonomy visualization'' of torsion.

Recalling that under parallel transport along $V{}^\mu$, the vector $U{}^\mu$ changes according to
\begin{align}
V{}^\alpha \partial{}_\alpha U{}^\mu = - \Gamma{}^\mu{}_{\alpha\beta} V{}^\alpha U{}^\beta \overset{*}{=} - K{}^\mu{}_{\alpha\beta} V{}^\alpha U{}^\beta \, ,
\end{align}
we can write the deficit around the closed circuits as
\begin{align}
\begin{split}
\label{eq:displacement}
\delta{}^\mu = -\delta\lambda^2 \Big[ 
&- V{}^\alpha K{}^\mu{}_{\alpha\beta}(x_1) U{}^\beta \\
&+ U{}^\alpha K{}^\mu{}_{\alpha\beta}(x_2) (U{}^\beta + V{}^\beta) \\
&+ V{}^\alpha K{}^\mu{}_{\alpha\beta}(x_3) V{}^\beta \Big] + \mathcal{O}(\delta\lambda^3) \, .
\end{split}
\end{align}
The scaling with $\delta\lambda^2$ originates from the fact that we follow the autoparallels only for a short amount of proper time $\delta\lambda$, and further supports the interpretation of this visualization technique as a holonomy since it approximately scales with the area enclosed by the circuit.

\begin{figure*}[!htb]
\centering
\includegraphics[width=0.32\textwidth]{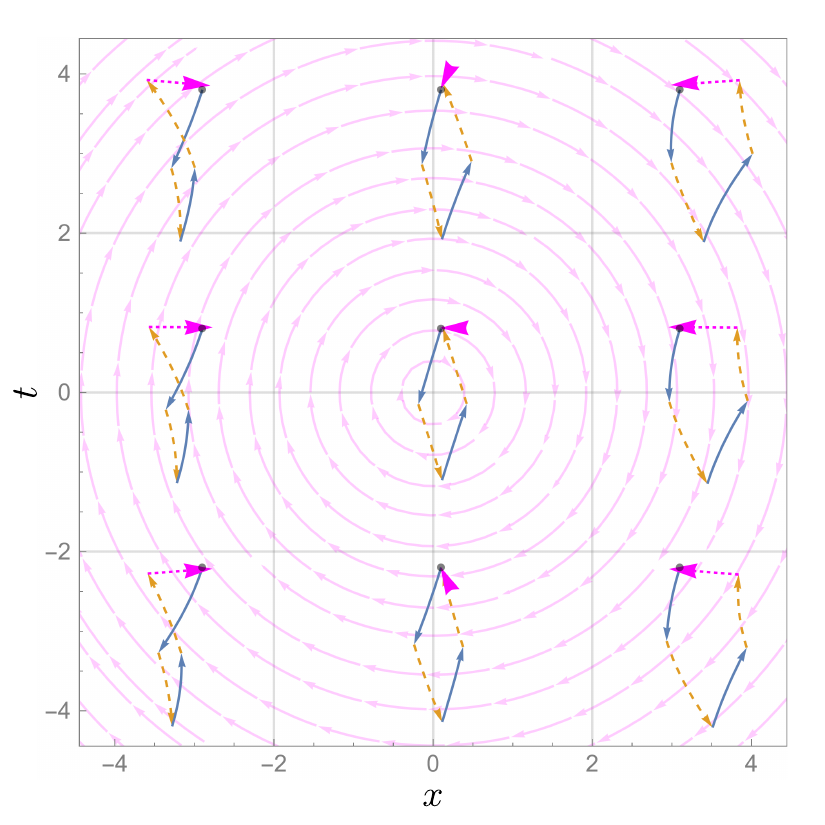}
\includegraphics[width=0.32\textwidth]{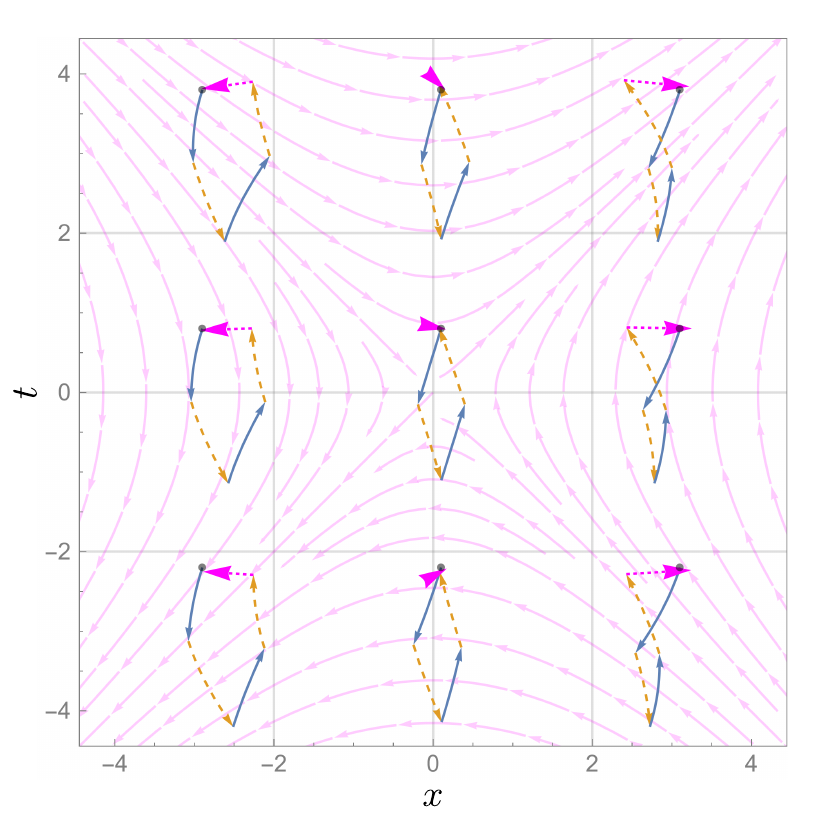}
\includegraphics[width=0.32\textwidth]{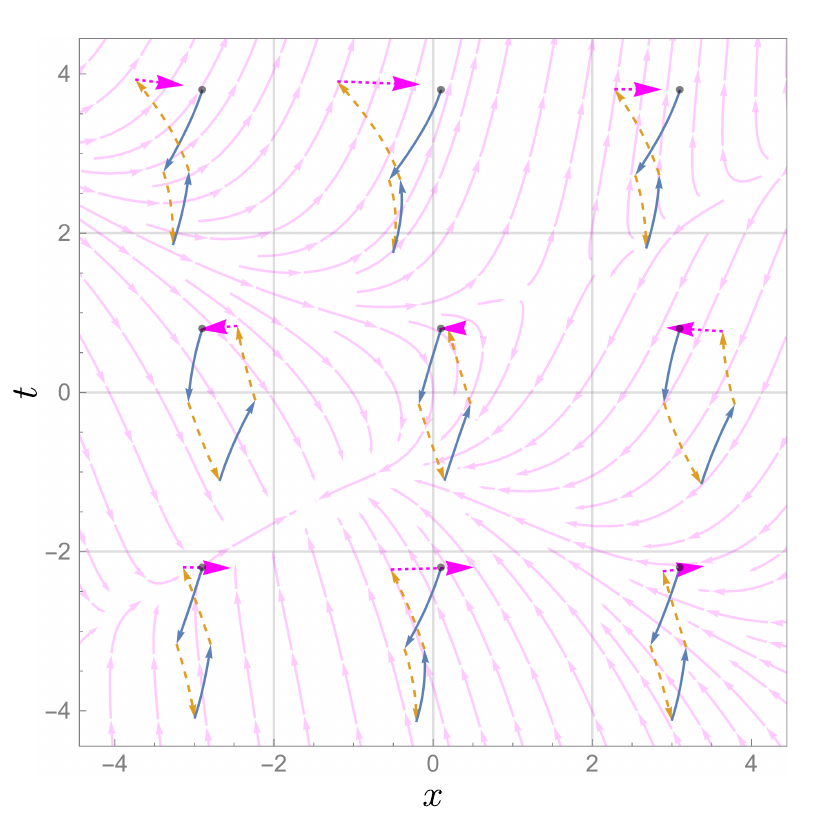}
\includegraphics[width=0.32\textwidth]{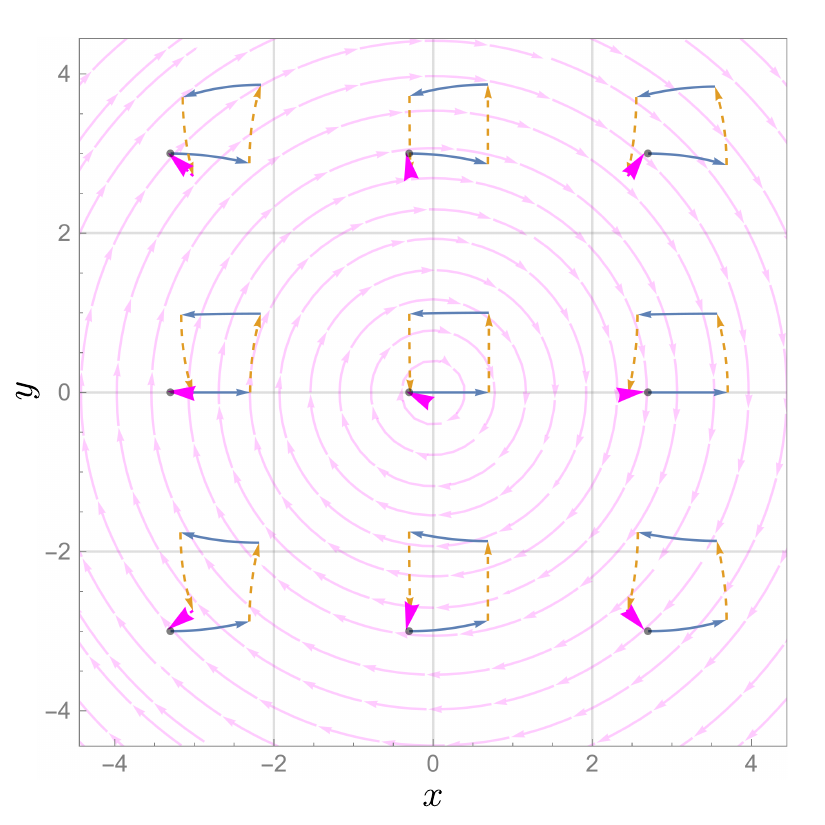}
\includegraphics[width=0.32\textwidth]{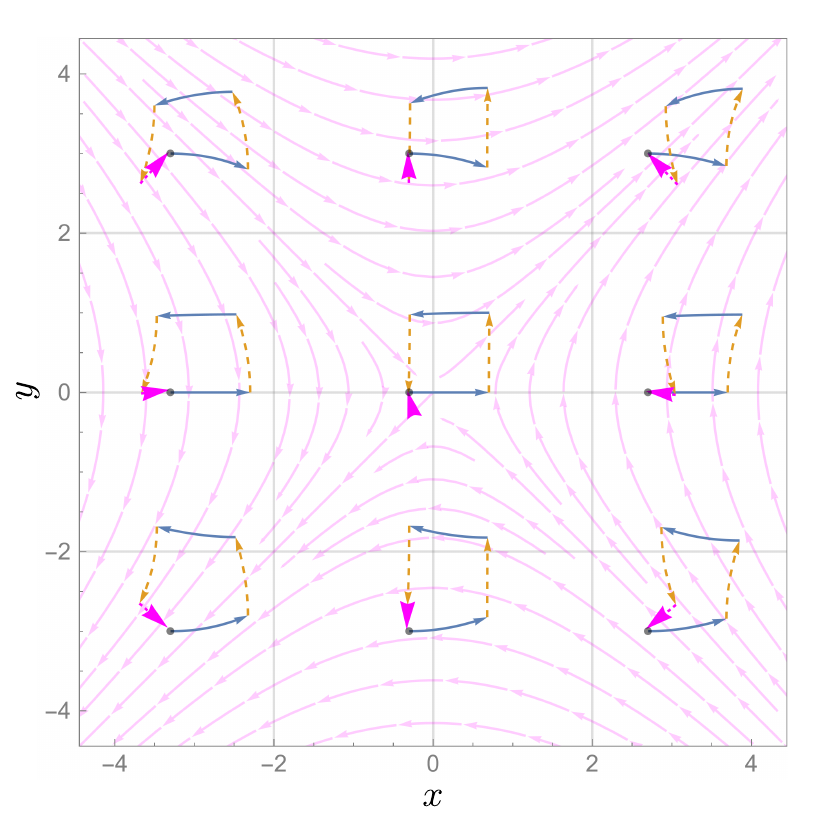}
\includegraphics[width=0.32\textwidth]{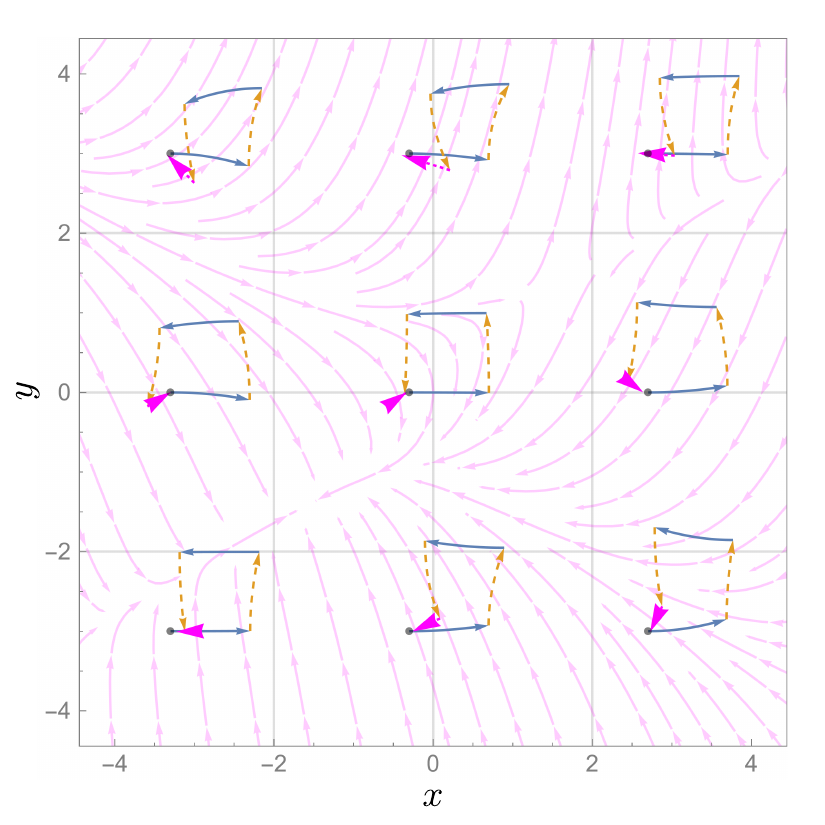}
\caption{We plot the holonomy visualization of the three torsion fields \eqref{eq:ex1}, \eqref{eq:ex2}, and \eqref{eq:ex3}. Left to right: scan over three torsion example configurations \eqref{eq:ex1}--\eqref{eq:ex3}; first row: spacetime case ($\epsilon=+1$), second row: spatial case ($\epsilon=-1$). For nine sample points in each diagram, we plot the corresponding circuit consisting of four autoparallels; autoparallels with antiparallel initial velocities are highlighted in the same color and linestyle; we moreover indicate the direction of the initial velocity by an arrowhead on the last point of each autoparallel. In the spacetime case the autoparallels' initial velocity is chosen to be $v=0.4$, and in the spatial case we choose unit velocity $v=1$. Moreover, we set $\delta\lambda=0.4$ throughout. Note that the circuits are rescaled by a factor of $(\delta\lambda)^{-1} = 2.5$ for enhanced visibility. While not in perfect agreement, the displacement vector is to good approximation given by the displacement vector computed in \eqref{eq:displacement}. }
\label{fig:holonomy-plots}
\end{figure*}

At the same time, the above expression is approximate in nature and only holds for autoparallels with a maximum proper time $\delta\lambda \ll 1$, making this method semi-local (that is, valid in a spacetime region of size $\delta\lambda$ around a fiducial point $x_0$). This means for practical purposes (we choose $\delta\lambda=0.4$) that the resulting circuits appear rather small when superimposed with the streamline plot of the background torsion field. For this reason we scale the circuit by the factor $\delta\lambda^{-1}$ when visualizing. For the above example torsion configurations \eqref{eq:ex1}--\eqref{eq:ex3} we plot the four scaled autoparallels with the corresponding scaled displacement vector in Fig.~\ref{fig:holonomy-plots}.

\subsection{Autoparallel visualization (semi-global)}

Moving away from the local notion of torsion and infinitesimal parallelograms and the semi-local notion of autoparallel circuits and their holonomies, we would now like to focus on global properties of torsion fields. Generically, without any additional assumptions on its symmetry properties, torsion in two-dimensional space(time) corresponds to a vector field. As such, in two-dimensional settings it can be visualized straightforwardly via streamline plots. However, this is not always satisfactory since the properties of autoparallel curves---and how their shape is affected by the presence of torsion---does not follow intuitively from the streamlines of the torsion vector field. Formulated in the two-dimensional setting, the autoparallel equation is
\begin{align}
\ddot{t} &= +(\epsilon\chi\dot{x} - \tau\dot{t}) \dot{x} \, , \\
\ddot{x} &= -(\epsilon\tau\dot{t} - \chi\dot{x}) \dot{t} \, .
\end{align}
The flat metric enters these expressions via the factor $\epsilon$ that stems from the raising and lowering of torsion indices in the contortion tensor, cf. Eq.~\eqref{eq:contortion}. If the expressions in parentheses were constant coefficients, the motion would correspond to that of a particle under influence of a Lorentz-type force, as one may expect in the presence of a vector field. However, since torsion is a 2-form-valued vector, it instead couples to a bilinear of velocities, somewhat obfuscating a physical intuition for autoparallel motion.

\begin{figure*}[!htb]
\centering
\includegraphics[width=0.32\textwidth]{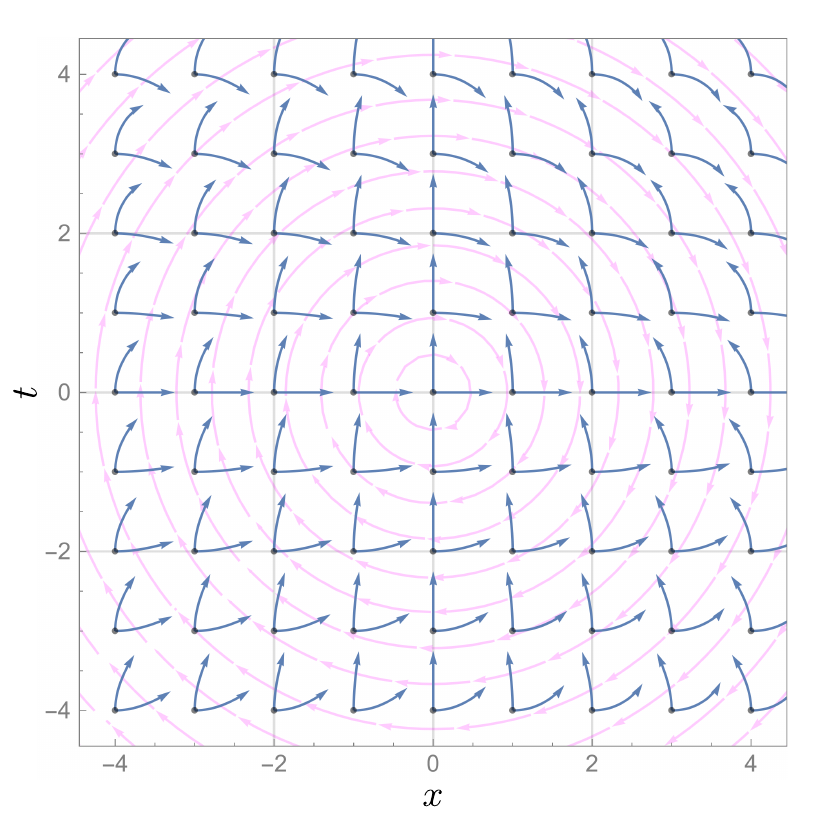}
\includegraphics[width=0.32\textwidth]{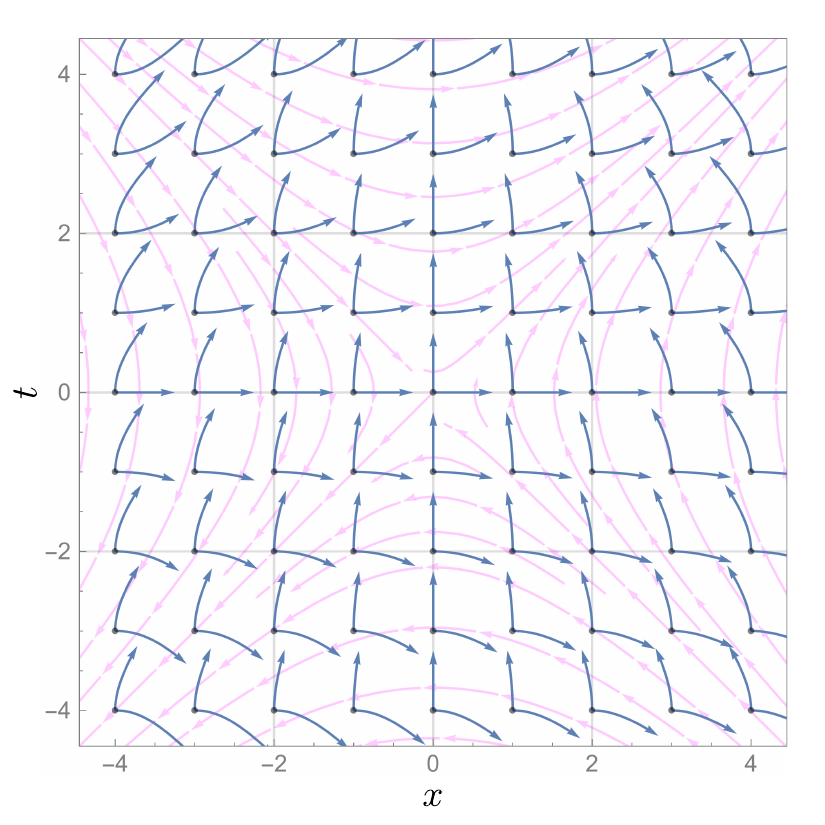}
\includegraphics[width=0.32\textwidth]{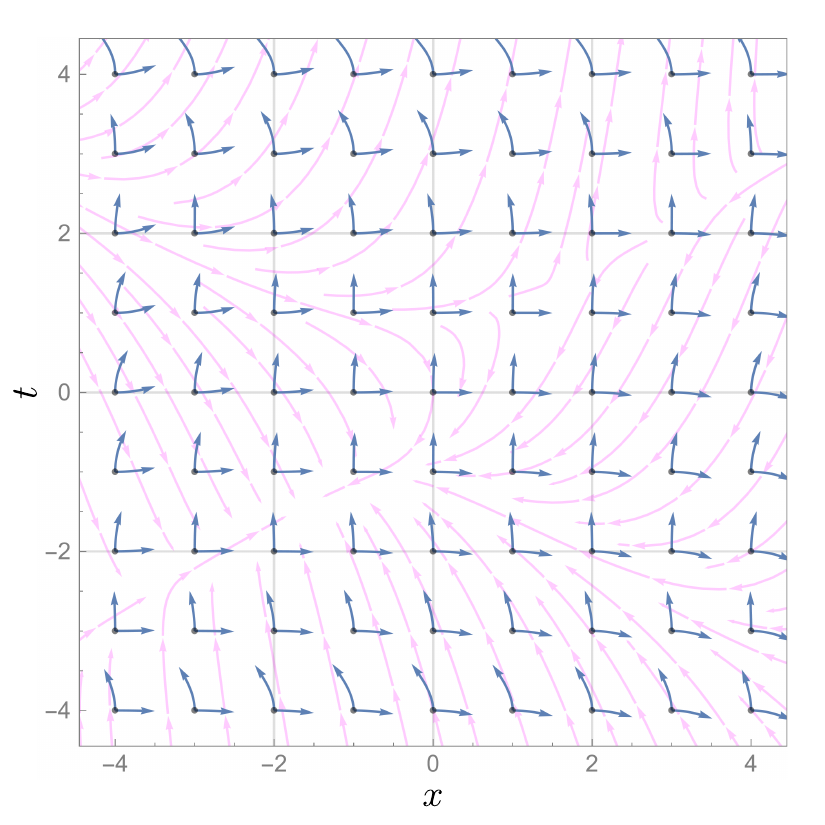}
\includegraphics[width=0.32\textwidth]{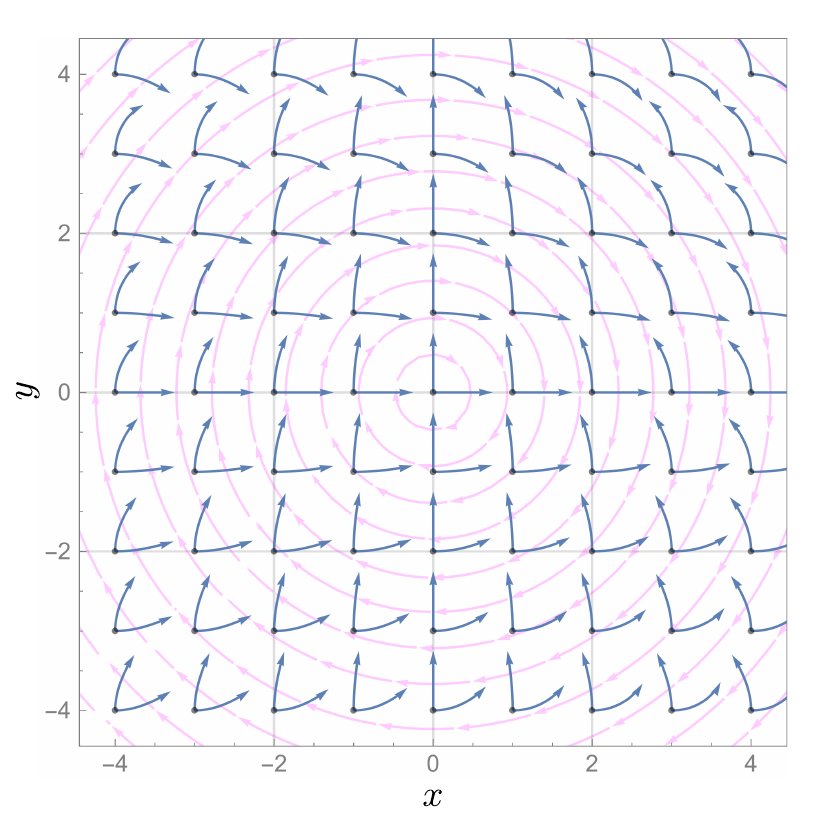}
\includegraphics[width=0.32\textwidth]{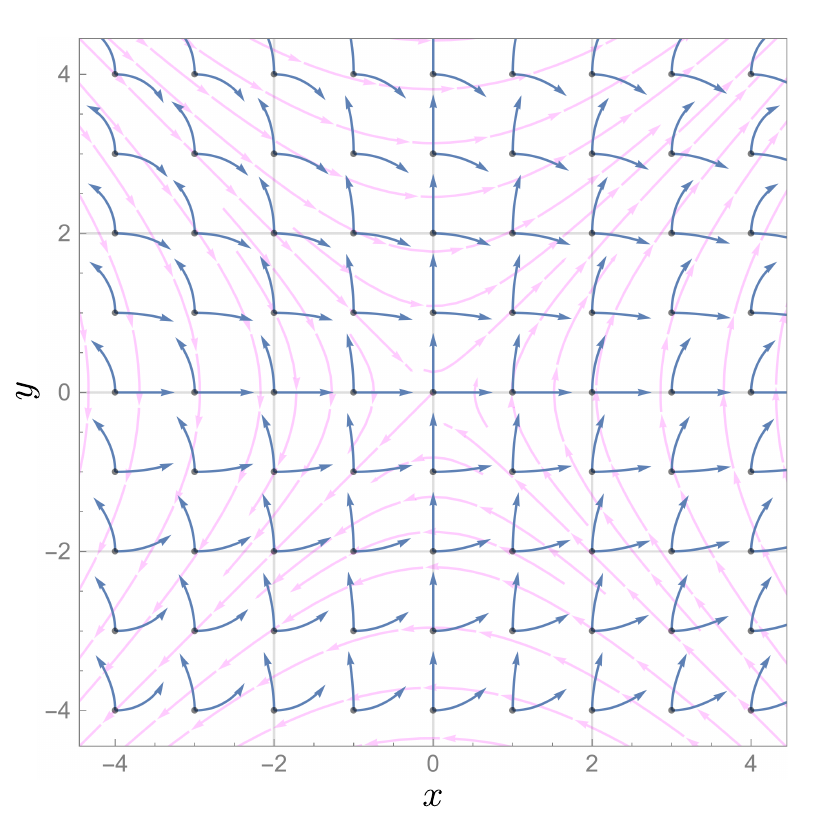}
\includegraphics[width=0.32\textwidth]{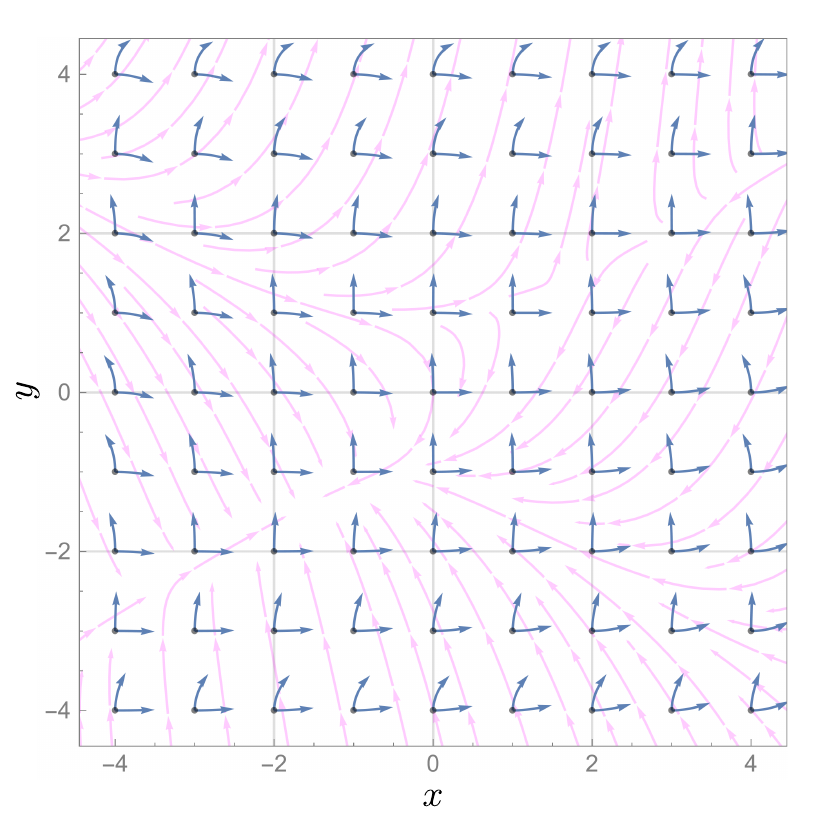}
\caption{We plot the holonomy visualization of the three torsion fields \eqref{eq:ex1}, \eqref{eq:ex2}, and \eqref{eq:ex3}. Left to right: scan over three torsion example configurations \eqref{eq:ex1}--\eqref{eq:ex3}; first row: spacetime case ($\epsilon=+1$), second row: spatial case ($\epsilon=-1$). For $9\times 9$ sample points in each diagram, we plot the corresponding autoparallel visualization.}
\label{fig:autoparallel-plots}
\end{figure*}

For this reason, we propose a visualization of torsion fields that superimposes the streamplot of the torsion background together with finite-range autoparallels. These autoparallels are defined to originate at each fiducial coordinate point in two arbitrarily chosen directions; however, since the background spacetime is flat, it is suggestive to take the vector fields $(1,0$ and $(0,1)$ as initial velocities. By comparing how the autoparallels' shapes change as one varies the fiducial coordinate point $x_0$, one is then able to infer properties of autoparallel motion, how it depends on the background torsion field, and how it changes semi-globally. We plot this for the three torsion example configurations in Fig.~\ref{fig:autoparallel-plots}.

This concludes the section on torsion visualizations in two space(time) dimensions. We close by pointing out that these methods can also be applied to higher-dimensional torsion fields of a certain symmetry (say, four-dimensional spherically symmetric fields) since one would be able to suppress spherical directions and focus solely on the temporal-radial sector. We will leave such investigations and possible further developments of the three visualization techniques to the future.

\section{Constant torsion: symmetries}
\label{sec:5}
From now on, we focus on constant torsion. Equation~\eqref{eq:curvature-in-terms-of-torsion} tells us that the curvature is hence zero, which simplifies the subsequent considerations. To that end, let us briefly set torsion to zero and recall the isometries of the resulting flat spacetime. It is encoded by the three Killing vectors
\begin{align}
\widetilde{A}{}^\mu = (1,~ 0) \, , \quad
\widetilde{B}{}^\mu = (0,~ 1) \, , \quad
\widetilde{C}{}^\mu = (x,~ \epsilon\, t) \, .
\end{align}
While the Killing vectors $\widetilde{A}{}^\mu$ and $\widetilde{B}{}^\mu$ encode translation invariance, the third Killing vector $\widetilde{C}{}^\mu$ is a result of boost invariance. This is also highlighted by the norms of these vectors:
\begin{align}
\widetilde{A}\cdot\widetilde{A} = -\epsilon \, , \quad
\widetilde{B}\cdot\widetilde{B} = 1 \, , \quad
\widetilde{C}\cdot\widetilde{C} = t^2 - \epsilon x^2 \, .
\end{align}
In the spacetime setting we have $\epsilon=1$ and hence $\widetilde{A}{}^\mu$ is timelike, corresponding to time translation invariance. Similarly, $\widetilde{C}{}^\mu$ becomes null on the surface $t^2=x^2$, implying that in the spacetime setting it corresponds to the boost Killing vector. These three Killing vectors satisfy the algebra
\begin{align}
[\widetilde{A}, \widetilde{B}] = 0 \, , \quad
[\widetilde{A}, \widetilde{C}] = \epsilon \widetilde{B} \, , \quad
[\widetilde{B}, \widetilde{C}] = \widetilde{A} \, ,
\end{align}
which is the Poincar\'e algebra in two-dimensional spacetime ($\epsilon=+1$) or two-dimensional space ($\epsilon = -1$). For any Killing vector $K{}^\mu$ one may compute the antisymmetric generator matrix $\nabla{}_\mu K{}_\nu$ that, when exponentiated, gives rise to an isometry transformation. In the present case, these generator matrices are given by
\begin{align}
\widetilde{\nabla}{}^\mu \widetilde{A}{}_\nu = 0 \, , \quad
\widetilde{\nabla}{}^\mu \widetilde{B}{}_\nu = 0 \, , \quad
\widetilde{\nabla}{}^\mu \widetilde{C}{}_\nu = \left( \begin{matrix} 0 & 1 \\ \epsilon & 0 \end{matrix} \right) \, .
\end{align}
The non-trivial, one-parametric $\widetilde{C}$-transformation is
\begin{align}
\hspace{-5pt} \exp\left[ \left( \begin{matrix} 0 & 1 \\ \epsilon & 0 \end{matrix} \right) \widetilde{\eta} \right]
= \left( \begin{matrix} \cosh(\sqrt{\epsilon}\widetilde{\eta}) & \frac{1}{\sqrt{\epsilon}} \sinh(\sqrt{\epsilon}\widetilde{\eta}) \\ \sqrt{\epsilon} \sinh(\sqrt{\epsilon}\widetilde{\eta}) & \cosh(\sqrt{\epsilon}\widetilde{\eta}) \end{matrix} \right) \, .
\end{align}
As expected, we recognize a boost of rapidity $\widetilde{\eta}$ for $\epsilon=+1$ and a rotation around the angle $\widetilde{\eta}$ for $\epsilon=-1$.

With these concepts briefly revisited, let us now study the consequences of turning on a constant torsion background,
\begin{align}
T{}_{tx}{}^t = \tau_0 \, , \quad T{}_{tx}{}^x = \chi_0 \, .
\end{align}
The autoparallel Killing equation for a Killing vector $\xi{}^\mu = (a(t,x), b(t,x))$ takes the form
\begin{align}
\begin{split}
\dot a + \tau b &= 0 \, , \\
b' - \chi a &= 0 \, , \\
\epsilon a' - \dot b - \epsilon\tau a - \chi b &= 0 \, .
\end{split}
\end{align}
Due to the nature of the equations, an exponential ansatz seems reasonable. We find the following solutions:
\begin{align}
A{}^\mu &= \left( \cosh s, ~ -\sqrt{\epsilon}\sinh s \right) \, , \\
B{}^\mu &= \left( -\sqrt{\epsilon}\sinh s, ~ \epsilon \cosh s \right) \, , \\
C{}^\mu &= \frac{1}{\epsilon \tau_0^2 - \chi_0^2} e^{\bar{s}} (\tau_0, ~\chi_0) \, ,
\end{align}
and their dependence on the coordinates $t$ and $x$ is given via the following quantities:
\begin{align}
s &\equiv \frac12 \epsilon{}^{\alpha\beta} T{}_{\alpha\beta\gamma} x{}^\gamma = \frac{\epsilon \tau_0 t - \chi_0 x}{\sqrt{\epsilon}} \, , \\
\bar{s} &\equiv T{}_{\alpha\beta}{}^\alpha x{}^\beta = - \chi_0 t + \tau_0 x \, .
\end{align}
This set of three autoparallel Killing vectors, according to a theorem by Peterson and Bonder \cite{Peterson:2019uzn}, is also maximal in two dimensions, since in $n$ spacetime dimensions there is a maximum of $n(n+1)/2$ autoparallel Killing vectors. Their norms are given by
\begin{align}
\hspace{-10pt}
A \cdot A = -\epsilon \, , \quad
B \cdot B = 1 \, , \quad
C \cdot C = (-1)\frac{e^{2\bar{s}}}{\epsilon\tau_0^2 - \chi_0^2} \, .
\end{align}
In the limit of vanishing torsion we find that
\begin{align}
\lim_{\tau,\chi \rightarrow 0} A{}^\mu = \widetilde{A}^\mu \, , \quad
\lim_{\tau,\chi \rightarrow 0} B{}^\mu = \widetilde{B}^\mu \, ,
\end{align}
whereas the limit of $C{}^\mu$ is singular. However, note that the three vectors $A{}^\mu$, $B{}^\mu$, and $C{}^\mu$ satisfy a deformed (but closed) Poincar\'e algebra,
\begin{align} \hspace{-5pt}
\label{eq:deformed-poincare-algebra-c}
[A, B]_\text{T} = 0 \, , \quad
[A, C]_\text{T} = \epsilon \, e^{\bar{s}} B \, , \quad
[B, C]_\text{T} = e^{\bar{s}} A \, .
\end{align}
which motivates the identification $C{}^\mu \leftrightarrow \widetilde{C}{}^\mu$ in the case of non-vanishing torsion. This is a central result of this paper, and we emphasize that the normalization of the vector $C$ is chosen such that the right-hand side of this algebra smoothly connects to the Poincar\'e algebra of flat spacetime. We moreover note that when utilizing the ordinary commutator $[\bullet,\bullet]$, the algebra of $A$, $B$, and $C$ would not close. Moreover, computing again the generators for the three autoparallel Killing vectors, we find
\begin{align}
\nabla{}^\mu A{}_\nu = 0 \, , \quad
\nabla{}^\mu B{}_\nu = 0 \, , \quad
\nabla{}^\mu C{}_\nu = e^{\bar{s}} \, \left( \begin{matrix} 0 & \epsilon \\ 1 & 0 \end{matrix} \right) \, .
\end{align}
The non-trivial, one-parametric $C$-transformation is
\begin{align}
\hspace{-5pt} \exp\left[ \left( \begin{matrix} 0 & \epsilon \\ 1 & 0 \end{matrix} \right) \eta \right]
&= \left( \begin{matrix} \cosh(\sqrt{\epsilon}\eta) & \sqrt{\epsilon} \sinh(\sqrt{\epsilon}\eta) \\ \frac{1}{\sqrt{\epsilon}}\sinh(\sqrt{\epsilon}\eta) & \cosh(\sqrt{\epsilon}\eta) \end{matrix} \right) \, , \\
\eta &= e^{\bar{s}} \, \widetilde{\eta} \, ,
\end{align}
where $\widetilde{\eta}$ is a free parameter, and the above corresponds to a spacetime-dependent Lorentz boost of rapidity $e^{\bar{s}} \eta_0$ (for $\epsilon = +1$) and to a space-dependent rotation of angle $e^{\bar{s}} \eta_0$ in the case of $\epsilon = -1$. In the limiting case of vanishing torsion, this generator matrix again reduces to the boost generator (or rotation generator) known from the torsionless case above. This further underscores the interpretation of $C{}^\mu$ as the torsionful generalization of the boost Killing vector.

However, a striking property of the flat spacetime boost Killing vector is its ability to achieve zero norm, which $C{}^\mu$ does not attain: depending on the sign of $\tau_0$ and $\chi_0$, it is either timelike everywhere or spacelike everywhere, and only assumes null values whenever $\bar{s} \rightarrow -\infty$.

For this reason, let us now briefly focus on the $\epsilon=1$ case and define the vector
\begin{align}
K = x A + \epsilon \, t B \, .
\end{align}
The vector $K{}^\mu$ is most decidedly \emph{not} an autoparallel Killing vector, but its algebra resembles a deformed Poincar\'e algebra,
\begin{align}
[A,K]_\text{T} = \left . B \right| _{T \rightarrow 2 T} \, , \\
[B,K]_\text{T} = \left . A \right| _{T \rightarrow 2 T} \, ,
\end{align}
where by ``$T \rightarrow 2T$'' we mean the substitution $\tau_0 \rightarrow 2\tau_0$ and $\chi_0 \rightarrow 2\chi_0$. Naturally, the norm of $K{}^\mu$ is that of a boost Killing vector,
\begin{align}
K \cdot K = t^2 - \epsilon x^2 \, .
\end{align}
Last, note that the autoparallel Killing equation for $K{}^\mu$ is proportional to $\sinh s$, implying that $K{}^\mu$ is a Killing vector provided
\begin{align}
\left. s\right|_{\epsilon = 1} = t\tau_0 - x\chi_0 = 0 \, .
\end{align}
While seeming far-fetched, we note that this condition is satisfied on the horizon of a two-dimensional black hole with null torsion. This brings us to the application of the above principles to static two-dimensional black holes in the presence of torsion (and, under certain assumptions, to four-dimensional black holes as well).

For completeness, we would like to close this section by some comments on the autoparallel equation of motion in constant torsion backgrounds, with a particular focus on conserved quantities. For the sake of simplicity, we focus here solely on the spacetime case $\epsilon=+1$. Parametrizing the 2-velocity of a particle as $u{}^\mu = (\dot{t}, \dot{x})$, we may write its normalization condition as ($\sigma=-1$ for a massive particle, $\sigma=0$ for a massless particle)
\begin{align}
g{}_{\mu\nu} u{}^\mu u{}^\nu = \sigma = - \dot{t}^2 + \dot{x}^2 \, .
\end{align}
The quantities $E = -g{}_{\mu\nu}A{}^\mu u{}^\nu$ and $P = +g{}_{\mu\nu}B{}^\mu u{}^\nu$ are the conserved energy and momentum, respectively. They take the explicit form
\begin{align}
E = \dot{t}\cosh s - \dot{x} \sinh s \, , \quad
P = \dot{x}\cosh s - \dot{t} \sinh s \, ,
\end{align}
which, when solved for $\dot{t}$ and $\dot{x}$, gives
\begin{align}
\dot{t} = E \cosh s + P \sinh s \, , \quad
\dot{x} = P \cosh s + E \sinh s \, .
\end{align}
Substituting this into the normalization condition then gives the dispersion relation
\begin{align}
\sigma = -E^2 + P^2 \, .
\end{align}
There also appears to be a third conserved quantity,
\begin{align}
Q = g{}_{\mu\nu} C^\mu u{}^\nu = \frac{1}{\tau_0^2 - \chi_0^2} e^{\bar{s}} (-\tau_0 \dot{t} + \chi_0 \dot{r})  \, .
\end{align}
While one may verify that this quantity is indeed constant under autoparallel motion for general $\tau_0$ and $\chi_0$, its physical interpretation remains elusive.

\section{Application to black holes}
\label{sec:6}

With the Killing vectors of flat spacetime and their counterparts in the presence of constant non-vanishing torsion introduced, let us now connect those topics to basic properties of static black holes in two (and, later, four) spacetime dimensions. This means we shall set
\begin{align}
\epsilon = 1
\end{align}
and exclusively consider the setting of spacetime. Static flat spacetime admits the existence of the hypersurface orthogonal timelike Killing vector $\widetilde{A}{}^\mu$ satisfying
\begin{align}
\widetilde{A}{}_{[\mu} \widetilde{\nabla}_\nu \widetilde{A}{}_{\rho]} = 0 \, .
\end{align}
Clearly, since $\nabla{}_\mu A{}_\nu = 0$, this definition of staticity carries over to the case of constant torsion. These are all ingredients required to track the influence of the presence of torsion onto the surface gravity of static, spherically symmetric two-dimensional black holes (and, additionally, to a subclass of four-dimensional ones).

\subsection{Two-dimensional static black holes}
Let us now, in this two-dimensional setting, imagine a black hole geometry. That is, for the moment we shall depart from the translational symmetry in the spatial direction, and consider the line element
\begin{align}
\dd s^2 = -f(r) \, \dd T^2 + \frac{\dd r^2}{f(r)} \, ,
\end{align}
where we denote the spatial direction by $r$ and label the time coordinate $T$ in order to distinguish them from the flat spacetime coordinates $x$ and $t$. Now, given a metric function $f(r)$ that asymptotes to unity at large distances but has a zero at finite radius $r_\text{h}$ and is positive throughout $r \in (r_\text{h}, \infty)$, we shall call this a black hole geometry. The metric function $f(r)$, close to the horizon $r_\text{h}$, hence admits the expansion
\begin{align}
f(r \approx r_\text{h}) = 2\widetilde{\kappa} (r-r_\text{h}) + \mathcal{O}\left[ (r-r_\text{h})^2 \right] \, ,
\end{align}
where the expansion coefficient $\widetilde{\kappa}$ is called the \emph{surface gravity} in the context of Riemannian geometry of general relativity,
\begin{align}
\widetilde{\kappa} \equiv \frac12 f'(r_\text{h}) \, .
\end{align}
Using methods of Euclidean gravity, it is related to the associated Hawking temperature $\widetilde{T}_\text{H}$ via
\begin{align}
\widetilde{T}_\text{H} = \frac{\widetilde{\kappa}}{2\pi} \, .
\end{align}
It is then useful to introduce a new set of coordinates that originates at the horizon. To that end, we define the near-horizon coordinate $z$ via
\begin{align}
r = r_\text{h} + \frac{\widetilde{\kappa}}{2} z^2 \, .
\end{align}
Then, near-horizon geometry is
\begin{align}
\dd s^2 \approx -\widetilde{\kappa}^2 z^2 \dd T^2 + \dd z^2 \, .
\end{align}
One recognizes flat spacetime under the Rindler coordinate transformation
\begin{align}
\label{eq:rindler-coordinates}
t = z \sinh(\widetilde{\kappa} T) \, , \quad
x = z \cosh(\widetilde{\kappa} T) \, ,
\end{align}
where the surface gravity $\widetilde{\kappa}$ of the static black hole plays the role of the constant acceleration parameter. The acceleration horizon is located at $z=0$, corresponding to the location of the black hole horizon in that approximation. Importantly, this implies that the properties of the near-horizon geometry of two-dimensional static black holes are inherited directly from the properties of flat spacetime. Moreover, the same result holds true in the presence of a non-trivial torsion background. This fact is guaranteed by the covariant nature of the autoparallel Killing equation. Hence, let us now connect the above results to the autoparallel Killing vectors, and, in particular, the boost generator.

Recalling the discussion from earlier, the boost generator of flat spacetime takes the following form in Rindler coordinates:
\begin{align}
\widetilde{K} = x\partial_t + t \partial_x = \frac{1}{\widetilde{\kappa}} \partial_T \, .
\end{align}
That is, up to a constant rescaling, the boost generator is responsible for time translations in the Rindler time coordinate $T$ \cite{Frolov:2011}. Introducing now outgoing ($u$) and ingoing ($v$) null coordinates in flat spacetime,
\begin{align}
u = \frac{t-x}{\sqrt{2}} \, , \quad
v = \frac{t+x}{\sqrt{2}} \, ,
\end{align}
the boost Killing vector $\widetilde{K}$ can be written as
\begin{align}
\widetilde{K} = -u\partial_u + v \partial_v \, .
\end{align}
This vector becomes null on the black hole horizon (where $t^2-x^2=0$). Since the horizon itself is a null surface, this vector can be interpreted as a generator of the horizon \cite{Frolov:2011,Poisson:2004}. In particular, on the future black hole horizon we have $u=0$ and can relate the Rindler time $T$ to the time coordinate $t$ of flat spacetime according to
\begin{align}
v \partial_v = \frac{1}{\widetilde{\kappa}} \partial_T \, .
\end{align}
Separating variables then leads to the relation
\begin{align}
\label{eq:kappa-def}
v = v_0 e^{\widetilde{\kappa}(T-T_0)} \, .
\end{align}
We may interpret this as follows: given the boost Killing vector $\widetilde{K}$ we can transform to null coordinates $u$ and $v$, consider the future horizon $u=0$, and extract the surface gravity $\widetilde{\kappa}$. Hence, the existence of a boost Killing vector is sufficient to derive the basic properties of the surface gravity of a static black hole. For this reason this presents us with a convenient avenue to define a notion of surface gravity in modified gravity theories purely through its isometries (and generalizations thereof).

In the context of the present work, we focus on the two vectors $C$ and $K$, which are the torsionful generalizations of the boost Killing vector $\widetilde{K}$ in a certain sense. Let us now explore if modified surface gravities can be defined from these objects.

\subsubsection{Surface gravity from the autoparallel Killing vector $C$}
The vector $C{}^\mu$ is a proper Killing vector, but does not become null at finite $t$ or $x$. Moreover, in the limit of vanishing torsion, its behavior is singular, and hence we expect that it does not give rise to a surface gravity that is in some sense a modified version of that from general relativity, but rather an ``additive'' quantity, in the sense that for zero torsion it approaches a universal constant value. We recall its form in $t$ and $x$ coordinates,
\begin{align}
C = \frac{1}{\tau_0^2 - \chi_0^2} e^{\tau_0 x - \chi_0 t} \left( \tau_0 \partial_t + \chi_0 \partial_x \right) \, ,
\end{align}
and now demand that this generates time translations in the Rindler coordinate $T$ as follows:
\begin{align}
C \equiv \frac{1}{\widetilde{\kappa}} \partial_T \, .
\end{align}
Converting to $u$ and $v$ coordinates, and setting $u=0$, we find on the future horizon that
\begin{align}
C = \frac{1}{\sqrt{2}} e^{-\Delta v_0} \left[ \frac{1}{\tau_0+\chi_0} \partial_u + \frac{1}{\tau_0-\chi_0} \partial_v \right]  \, ,
\end{align}
where for subsequent simplicity we defined
\begin{align}
\Delta \equiv \frac{\tau_0-\chi_0}{\sqrt{2}} \, .
\end{align}
Integrating (and ignoring the $\partial_u$-terms), one extracts
\begin{align}
v(T) = - \frac{1}{\Delta} \ln \left[ e^{- \Delta v_0} - \frac{\widetilde{\kappa}}{2}(T-T_0) \right ] \, .
\end{align}
In analogy to Eq.~\eqref{eq:kappa-def} we now define a torsionful surface gravity, derived from the vector field $C$, as the quantity
\begin{align}
\kappa_C \equiv \left. \frac{\partial v(T)}{\partial T}\right|_{\substack{T=T_0\\v_0=1}} = \frac{\widetilde{\kappa}\,e^{- \Delta}}{2\Delta} \, .
\end{align}
The corresponding Hawking temperature then is
\begin{align}
T_C \equiv \frac{\kappa_C}{2\pi} = \frac{\widetilde{\kappa}\,e^{- \Delta}}{4\pi\Delta} \, .
\end{align}
The limit of these quantities as torsion vanishes (that is, as $\Delta \rightarrow 0$) is undefined, stemming from the normalization of the vector $C$ itself. Note, however, that this normalization of the vector was necessitated by smoothly connecting the right hand-side of the deformed Poincar\'e algebra \eqref{eq:deformed-poincare-algebra-c} to the flat spacetime limit.

\subsubsection{Approximate surface gravity from the null vector $K$}

Unlike $C{}^\mu$, the vector $K^\mu$ is \emph{not} a Killing vector. However, in the case of null torsion, that is,
\begin{align}
\tau_0 = \chi_0 \quad \Rightarrow \quad T{}_{\mu\nu\rho} T{}^{\mu\nu\rho} = 0 \, , 
\end{align}
it becomes a Killing vector on the horizon, where it also becomes null. Moreover, in the limit of vanishing torsion it smoothly approaches the boost Killing vector $\tilde{K}{}^\mu$. We may therefore expect a smooth expression for the surface gravity derived from $K{}^\mu$ that reduces to the general relativistic expression in the limit of vanishing torsion, resembling a multiplicative modification. However, on the horizon we simply find that
\begin{align}
K = v \, e^{-\Delta v} \partial_v = v \partial_v \, ,
\end{align}
since the null torsion assumption immediately sets $\Delta = 0$, and we hence arrive at the identical expression encountered in general relativity.

Let us briefly consider the case $\Delta > 0$ but small, and compute the implications of setting
\begin{align}
K = v \, e^{-\Delta v} \partial_v = \frac{1}{\widetilde{\kappa}}\partial_T \, .
\end{align}
Integrating on both sides, one finds
\begin{align}
\widetilde{\kappa}(T-T_0) = \text{Ei}\left[\Delta v \right] - \text{Ei}\left[ \Delta v_0 \right] \, .
\end{align}
It is clear that for $\tau_0 = \chi_0 = 0$ one recovers the standard relation \eqref{eq:kappa-def} from above, since in that limit $\Delta = 0$ and hence
\begin{align}
\text{Ei}\left[\Delta v \right] - \text{Ei}\left[ \Delta v_0 \right] = \ln \frac{v}{v_0} \, .
\end{align}
For small torsion coefficients, $(\tau_0-\chi_0)v \ll 1$ and $(\tau_0-\chi_0)v_0 \ll 1$, one finds
\begin{align}
\widetilde{\kappa}(T-T_0) \approx \ln\left( \frac{v}{v_0} \right) + \Delta (v-v_0)
\end{align}
This can be solved, at least formally, in terms of the Lambert-W function,
\begin{align}
v(T) = \frac{1}{\Delta} W \left[ \Delta v_0 \, e^{\Delta v_0 + \widetilde{\kappa}(T-T_0)} \right]
\end{align}
We now define the approximate surface gravity $\kappa_\text{K}$ as
\begin{align}
\kappa_\text{K} \equiv \left. \frac{\partial v(T)}{\partial T}\right|_{\substack{T=T_0\\v_0=1}} = \frac{\widetilde{\kappa}}{\Delta} \frac{ W\left(\Delta \, e^{\Delta} \right) }{ 1 + W\left(\Delta \, e^{\Delta} \right) } \, .
\end{align}
Then, for small torsion, we can extract
\begin{align}
\kappa_\text{K} = \widetilde{\kappa} \left( 1 - \Delta \right) \, .
\end{align}
For vanishing torsion we recover $\kappa_\text{K} = \widetilde{\kappa}$, as required for consistency.

\subsubsection{Comparison}
\begin{figure}[!htb]
\centering
\includegraphics[width=0.48\textwidth]{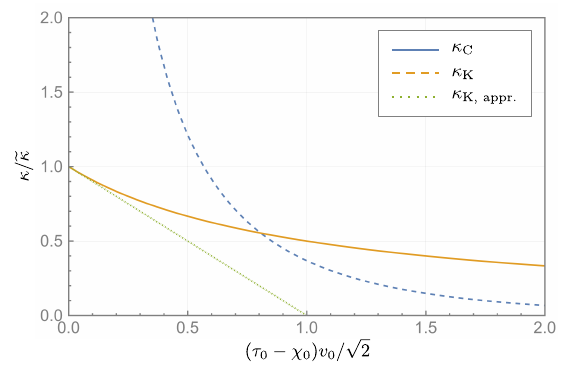}
\caption{We plot the different notions of surface gravities ($\kappa_\text{C}$, $\kappa_\text{K}$, and the approximated form of $\kappa_\text{K}$) as a function of the dimensionless quantity $\Delta v_0$, normalized to the Riemannian surface gravity $\widetilde{\kappa}$.}
\label{fig:surface-gravities}
\end{figure}

Let us briefly compare the autoparallel surface gravity $\kappa_\text{C}$ and the approximate surface gravity $\kappa_\text{K}$, given by
\begin{align}
\frac{\kappa_\text{C}}{\widetilde{\kappa}} &= \frac{e^{- \Delta}}{2\Delta} \, , \\
\frac{\kappa_\text{K}}{\widetilde{\kappa}} &= \frac{1}{\Delta} \frac{ W\left(\Delta \, e^{\Delta} \right) }{ 1 + W\left(\Delta \, e^{\Delta} \right) } \approx 1 - \Delta \, .
\end{align}
Notably, these quantities do not bring about an additional dependence on the Riemannian surface gravity $\widetilde{\kappa}$ and solely depend on the torsion coefficients $\tau_0$ and $\chi_0$ via $\Delta = (\tau_0-\chi_0)/\sqrt{2}$. However, since the entire discussion has been performed irrespective of the field equations of the two-dimensional gravitational model with torsion, it remains a possibility that the torsion coefficients depend on the mass parameter $M$ of the black hole. This, in turn, would then lead to interesting modifications of the thermodynamical properties of such geometries, which could allow for a direct comparison to results presented elsewhere \cite{Dey:2017fld}. However, such comparative studies are beyond the scope of the present work, but may be addressed in future work. See Fig.~\ref{fig:surface-gravities} for a visualization.

\subsection{Four-dimensional static and spherically symmetric black holes}

In order to generalize the preceding notions of surface gravities in the presence of torsion to spherically symmetric and static black hole geometries in four dimensions, we emphasize that such static spherical geometries generally factorize into temporal-radial part as well as a spherical part,
\begin{align}
\begin{split}
\dd s^2 &= -a(r) \, \dd t^2 + b(r) \dd r^2 + r^2 \dd\Omega^2 \, , \\
\dd\Omega^2 &= \dd\theta^2 + \sin^2\theta\,\dd\varphi^2 \, ,
\end{split}
\end{align}
where $a(r)$ and $b(r)$ are independent functions that only depend on the radial coordinate due to the assumed staticity of the geometry.

In spacetimes respecting spherical symmetry, the following torsion components are non-vanishing \cite{Damour:2019oru}:
\begin{align}
T{}_{tr}{}^t \, , \quad
T{}_{tr}{}^r \, , \quad
T{}_{t\theta}{}^\theta \, , \quad
T{}_{t\varphi}{}^\varphi \, , \quad
T{}_{r\theta}{}^\theta \, , \quad
T{}_{r\varphi}{}^\varphi \, .
\end{align}
Importantly, as long as the spherical part and the radial part decouple (in a suitable sense to be discussed below), the presence of a spherically symmetric submanifold is compatible with the existence of other Killing vectors generated in the temporal-radial sector. For example, in the case of the Schwarzschild geometry, the vector $\partial_t$ is always a Killing vector, even when the Schwarzschild geometry is lifted to higher dimensions.

Setting now $a(r) = 1/b(r)$ and introducing again the near-horizon coordinate $r = r_\text{h} + \frac{\widetilde{\kappa}}{2} z^2$, the four-dimensional near-horizon geometry is given by
\begin{align}
\begin{split}
\dd s^2 &\approx -\widetilde{\kappa}^2 z^2 \dd T^2 + \dd z^2 + r_h^2 \dd\Omega^2 \\
&= - \dd t^2 + \dd x^2 + r_h^2 \dd\Omega^2 \, ,
\end{split}
\end{align}
where the second equality follows from the Rindler coordinate transformation similar to \eqref{eq:rindler-coordinates},
\begin{align}
\label{eq:rindler-coordinates}
t = z \sinh(\widetilde{\kappa} T) \, , \quad
r = z \cosh(\widetilde{\kappa} T) \, .
\end{align}
The presence of the spherical sector implies that the near-horizon geometry is no longer flat, since its Riemannian (torsion-free) scalar curvature is
\begin{align}
\tilde{R} = \frac{2}{r_0^2} \, ,
\end{align}
generated by the non-trivial Christoffel symbols
\begin{align}
\widetilde{\Gamma}{}^\theta{}_{\varphi\varphi} = -\cos\theta\sin\theta \, , \quad
\widetilde{\Gamma}{}^\varphi{}_{\theta\varphi} = \widetilde{\Gamma}{}^\varphi{}_{\varphi\theta} = \cot\theta \, .
\end{align}
Putting this all together, we may now ask: can the two-dimensional autoparallel Killing vectors be lifted into four dimensions as solutions to the four-dimensional autoparallel Killing equation? In particular, given an initial two-dimensional autoparallel Killing vector expressed in two-dimensional coordinates $t$ and $x$,
\begin{align}
\xi_\text{2D}^\mu &= (\xi_\text{2D}^t, \xi_\text{2D}^x) \, ,
\end{align}
is the lifted quantity
\begin{align}
\xi_\text{4D}^\mu &\equiv \left( \xi_\text{2D}^t, \xi_\text{2D}^x, 0, 0 \right) \, ,
\end{align}
where we are identifying the $x$-component of the two-dimensional geometry with the $r$-component of the four-dimensional geometry, a solution of the autoparallel Killing equation?

First, the non-trivial Christoffel symbols do not mix the spherical $\theta\varphi$-sector with the temporal radial $tr$-sector, since they are purely non-zero in the angular part. This means that all Killing vectors in 2D are also Killing vectors in the lifted version. Second, however, the presence of torsion in the autoparallel Killing equation may spoil this behavior. In order to track the difference between the Killing equation and the autoparallel Killing equation we define the symmetric difference tensor
\begin{align}
\begin{split}
K{}_{\mu\nu} &= K{}_{\nu\mu} = \nabla{}_\mu \xi{}_\nu + \nabla{}_\nu \xi{}_\mu - \left( \widetilde{\nabla}_\mu \xi{}_\nu + \widetilde{\nabla
}_\nu \xi{}_\mu \right) \\
&= - 2K{}^\alpha{}_{(\mu\nu)} \xi{}_\alpha = \left( T{}_\mu{}^\alpha{}_\nu + T{}_\nu{}^\alpha{}_\mu \right) \xi{}_\alpha \, .
\end{split}
\end{align}
Hence, non-trivial torsion coefficients can mix components of the temporal-radial $tr$-sector with the angular $\theta\varphi$-sector. 
For a vector $\xi{}^\mu = (\xi^t, \xi^r, \xi^\theta, \xi^\varphi)$ we find
\begin{align}
\begin{split}
K_{tt} &= -2T{}_{tr}{}^t a(r) \xi^r \, , \\
K_{tr} &=  T{}_{tr}{}^t a(r) \xi^t + T{}_{tr}{}^r b(r) \xi^r \, , \\
K_{t\theta} &= r^2T{}_{t\theta}{}^\theta \xi^\theta \, , \\
K_{t\varphi} &= r^2\sin^2\theta \, T{}_{t\varphi}{}^\varphi \xi^\varphi \, , \\
K_{rr} &= -2T{}_{tr}{}^r b(r) \xi^t \, , \\
K_{r\theta} &= r^2 T{}_{r\theta}{}^\theta \xi^\theta \, , \\
K_{r\varphi} &= r^2\sin^2\theta \, T{}_{r\varphi}{}^\varphi \xi^\varphi \, , \\
K_{\theta\theta} &= -2r^2 \left( T{}_{t\theta}{}^\theta\xi^t + T_{r\theta}{}^\theta \xi^r \right) \, , \\
K_{\theta\varphi} &= 0 \, , \\
K_{\varphi\varphi} &= -2r^2\sin^2\theta \left( T{}_{t\varphi}{}^\varphi\xi^t + T_{r\varphi}{}^\varphi \xi^r \right) \, .
\end{split}
\end{align}
In order to test the decoupling condition we set the angular components to zero, $\xi^\theta = \xi^\varphi = 0$. One finds identical equations to the two-dimensional case, except in the $\theta\theta$ and $\varphi\varphi$ components. This means that the presence of torsion generally spoils the clear factorization of the temporal-radial sector and the spherical sector! The conditions for such a decoupling are therefore
\begin{align}
T{}_{t\theta}{}^\theta \xi{}^t + T{}_{r\theta{}}{}^\theta \xi{}^r = 0 \, , \quad
T{}_{t\varphi}{}^\varphi \xi{}^t + T{}_{r\varphi}{}^\varphi \xi{}^r = 0 \, .
\end{align}
A sufficient condition for this is
\begin{align}
T{}_{t\theta}{}^\theta = T{}_{t\varphi}{}^\varphi = T{}_{r\theta}{}^\theta = T{}_{r\varphi}{}^\varphi = 0 \, ,
\end{align}
with only $T_{tr}{}^t$ and $T{}_{tr}{}^r$ non-vanishing.

Then, for this class of spacetimes, the results of the two-dimensional section can directly be lifted to four dimensions. One may wonder if these restrictions are physical or not. While this particular case is for example \emph{too restrictive} to account for field configurations encountered in torsion bigravity \cite{Damour:2019oru} or the Baekler solution of quadratic Poincar\'e gauge gravity \cite{Baekler:1981lkh,Obukhov:2019fti,Obukhov:2022khx}, a special case that also has $T{}_{tr}{}^r = 0$ is considered by Peterson and Bonder \cite{Peterson:2019uzn} [see after their Eq.~(16)]. Other spherically symmetric configurations that do not satisfy these conditions have been considered by Sharif and Majeed \cite{Sharif:2009vz}.

\section{Conclusions}
\label{sec:7}

In this paper we studied two-dimensional torsion configurations, from an off-shell perspective without invoking field equations of a particular gravitational model. We developed local, semi-local, and semi-global visualization techniques of arbitrary two-dimensional torsion fields in both purely spatial and spacetime contexts, and then focused on the special case of constant torsion coefficients. There, we found a maximal set of three autoparallel Killing vectors and showed that they satisfy a deformed Poincar\'e algebra under the torsionful commutator $[\bullet,\bullet]_\text{T}$. Motivated by the emergence of such a structure, we extracted the generalized boost generator from this algebra and used it to define two notions of surface gravity in the presence of torsion in two spacetime dimensions. Finally, we studied under what conditions the two-dimensional results can be lifted to four-dimensional spacetime. As it turns out, these conditions on the torsion coefficients are rather restrictive and exclude several known black hole configurations encountered in the literature, but the plethora of different gravitational models, especially in the teleparallel sector, may prove surprising in the future.

Besides the technical details, a central outcome of this work is the following: in the presence of torsion, it is possible to define a notion of an autoparallel Killing vector $\xi{}^\mu$ that solves
\begin{align}
\nabla{}_\mu \xi{}_\nu + \nabla{}_\nu \xi{}_\mu = 0 \, .
\end{align}
While at face value this has nothing to do with isometries (which are always defined in terms of the Lie derivative, and can be suitable generalized to the case of non-vanishing torsion \cite{Obukhov:2022khx}), the set of autoparallel Killing vectors found in this paper still satisfies a deformed Poincar\'e algebra. This seemingly connects those autoparallel Killing vectors to fundamental isometric properties of the underlying spacetime. It remains to be seen if (and how) these results can be generalized to other scenarios of physical interest in the strong-gravity regime.

\section{Acknowledgements}

I would like to thank Helen Meskhidze (Harvard BHI) for discussions during the early stages of this work, and Friedrich W. Hehl (Cologne) for many years of correspondence on the topic of Poincar\'e gauge gravity and the physical properties of torsion. I am grateful for support as a Fellow of the Young Investigator Group Preparation Program, funded jointly via the University of Excellence strategic fund at the Karlsruhe Institute of Technology (administered by the federal government of Germany) and the Ministry of Science, Research and Arts of Baden-W\"urttemberg (Germany).

\end{document}